# Manganese oxide coated TiO$_2$ nanotube-based electrode for efficient and selective electrocatalytic sulfide oxidation to colloidal sulfur

*Natalia Sergienko[a,b], Jelena Radjenovic [a,c*]*

[a]*Catalan Institute for Water Research (ICRA), Scientific and Technological Park of the University of Girona, 17003 Girona, Spain*

[b] *University of Girona, Girona, Spain*

[c]*Catalan Institution for Research and Advanced Studies (ICREA), Passeig Lluís Companys 23, 08010 Barcelona, Spain*

*\* Corresponding author:*

*Jelena Radjenovic, Catalan Institute for Water Research (ICRA), c/Emili Grahit, 101, 17003 Girona, Spain*

Phone: + 34 972 18 33 80; Fax: +34 972 18 32 48; E-mail: jradjenovic@icra.cat

# Abstract

Manganese oxide-coated $TiO_2$ nanotube array (NTA) was synthesized and applied for the (electro)catalytic oxidation of sulfide. By setting the electrodeposition parameters and calcination temperature, the coating obtained was $MnO_2$ or $Mn_2O_3$, with rod or needle like morphology. Excellent catalytic activity of the Ti/$TiO_2$ NTA/$Mn_xO_y$ anodes resulted in robust and selective oxidation of sulfide to elemental sulfur via an inner sphere complex between the sulfide ion and Mn-oxide. The penetration of the $Mn_xO_y$ inside the NTA improved adhesion and enhanced the electron transfer, thus enabling complete and rapid electrocatalytic re-oxidation of the Mn-oxide coating. At pH 8, the final product of sulfide oxidation was colloidal sulfur, $S_8$, which prevented anode passivation and enabled a stable (electro)catalytic activity in the subsequent applications. This was observed in both synthetic electrolyte and real sewage, thus demonstrating the efficiency of the Ti/$TiO_2$ NTA/$Mn_xO_y$ anode for sulfide oxidation in complex waste streams.

**Keywords:** sulfide oxidation, colloidal sulfur, manganese oxide, electrocatalysis.

# Introduction

Sulfur is one of the most abundant elements on earth [1]. The pre-industrial sulfur cycle was modified by the increased anthropogenic activity, leading to numerous environmental problems. One of the major contributors to the disturbances of global sulfur balance is the urban water cycle. Wastewater, introduced into the municipal collection systems, is characterized by the abundance of sulfate, whose primary source is household cleaning detergents and alum-based flocculants used for potable water purification [2, 3]. Waste streams, rich in sulfate and organic matter, flowing through pipelines at low velocity, create particularly favorable conditions for the sulfate reducing bacteria (SRB) [4, 5]. As a result of the SRB activity, sulfate is rapidly reduced to sulfide, followed by the hydrogen sulfide release into the gaseous phase of the sewer pipe [6]. Even at concentration as low as 1-2 mg $L^{-1}$, hydrogen sulfide causes "rotten egg" odor typically associated with the sewer systems [7]. Hydrogen sulfide gas is heavier than air and tends to accumulate in sewers, reaching life threatening concentrations for people involved in its maintenance [8, 9]. Under certain biogenic pathway, hydrogen sulfide gas is oxidized into sulfuric acid, which attacks concrete and metal pipes, leading to premature deterioration of the sewer assets [10, 11]. Since sewers are placed underground, it is difficult to determine the level of corrosion, and as a result, this problem is often underestimated and overlooked. The cost of this negligence is high; for instance, Spain spends 10 million euros annually on the replacement of the sewer pipeline due to sulfide induced corrosion [12]. Another reason why sulfide control in sewers is not so common, especially at small-scale, are numerous drawbacks associated with the technologies currently available at the market. For example, physicochemical removal of sulfide requires continuous dosage of chemicals (e.g., nitrate, oxygen, caustic, lime, magnesium hydroxide, iron salts) and significant energy inputs, leading to high operational costs [13].

Moreover, such methods offer only temporary sulfide mitigation or lead to the production of secondary waste streams [13, 14].

Given the limited technical solutions and the large scale of the problem of sewer corrosion, in recent years researchers have turned their focus towards the electrochemical sulfide control [13, 15, 16]. The electrochemical sulfide control can be performed *in-situ*, thus eliminating all the risks and costs associated with the chemicals dosing, storage and transportation [13]. Moreover, electrochemical approach is selective and can avoid undesirable side-reactions [17].

Studies focused on the electrochemical removal of sulfide are mainly based on its non-selective oxidation to sulfate and thiosulfate by the electrogenerated oxygen [18, 19]. Selective oxidation of sulfide to sulfur can be achieved at carbon-based electrodes (e.g., graphite, carbon, activated carbon), thus allowing its complete separation from wastewater [15, 16, 20, 21]. Nevertheless, chemisorption of sulfide and its incorporation into the carbon matrix eventually leads to process failure due to electrode passivation with elemental sulfur, and limits the practical application of these materials [16].

Sulfide oxidation with manganese is a well-known process naturally occurring in sediments, yet it was barely exploited for sulfide removal from wastewater [22, 23]. The ability of Mn to switch reversibly between different redox states enables it to participate in numerous reactions, including charge rearrangement between two or more Mn ions [24, 25]. The high redox activity, earth abundance and low toxicity make Mn a perfect catalyst for environmental applications [26]. Indeed, Mn was successfully applied for the removal of various refractory organic contaminants, including aniline [27], formaldehyde [28, 29], phenol [30-32], as well as volatile organic contaminants [33, 34]. Several studies proposed the removal of sulfide through dosing of potassium permanganate [35, 36]. Even though demonstrating fast sulfide removal rates at first, the treatment performance gradually decreases as Mn becomes depleted due to its interaction with sulfide [37]. In our previous study, we demonstrated that application of low anodic potentials enabled continuous regeneration of the $Mn_xO_y$ catalyst coating, thus preventing its gradual depletion [38]. The $Mn_2O_3$-coated graphite felt (GF) electrodes rapidly oxidized sulfide to elemental sulfur that remained deposited at the anode surface [38]. Yet, an attempt to regenerate the GF/$Mn_2O_3$ anode through cathodic dissolution of the electrodeposited sulfur was only partially successful, as it also caused the reductive dissolution of the $Mn_2O_3$ coating.

To address the previously encountered limitations, in this study we developed a Ti plate anode with $Mn_xO_y$ coated $TiO_2$ nanotube array (NTA) interlayer, for rapid and selective oxidation of sulfide to elemental sulfur. The highly conductive Ti/$TiO_2$NTA is characterized by the high specific surface area, which enables the deposition of a larger amount of catalyst and increases the electrocatalytic activity of the electrode [39]. Moreover, it can improve adhesion and enhance the electron transfer between the coating and the substrate, thus ensuring a rapid recovery of the $Mn_xO_y$ coating [30, 40, 41]. We have investigated the impact of the operating parameters (i.e., applied potential, pH, sulfide concentration) on the sulfide removal kinetics and the formed reaction products. Furthermore, the synthesized Ti/$TiO_2$NTA-$Mn_xO_y$ electrodes were applied for the sulfide oxidation in real sewage to investigate the performance of the electrodes under realistic environmental conditions.

## 2. Materials and methods

### *2.1 Material synthesis*

#### *2.1.1 Synthesis of the TiO$_2$ nanotube array (NTA)*

4 × 4 cm Ti plates (1 mm, 99.6% purity, Advent Research Materials, U.K.) were mechanically polished until mirror finish. Polished materials were then degreased by sonication in isopropanol, acetone and methanol, rinsed with deionized water and dried in a nitrogen stream. Prior to the anodization, titanium plates were etched in 17% w/w HCl aqueous solution (Scharlab, Spain) at 75°C for 15 min to obtain fresh metal surface for the nanotubes growth. Next, the Ti plates were anodized in a mixture of glycerol and deionized water (50:50 vol.%) with 0.5 wt.% NH$_4$F, using a two electrode cell configuration and stainless steel mesh as the counter electrode. The potential of the cell was controlled with Autolab 302N potentiostat/galvanostat equipped with the voltage amplifier (Metrohm Autolab B.V., The Netherlands). The anodization procedure included ramping of the potential from the open circuit (OC) to 20 V and maintaining the potential at 20 V during 2 h. After the treatment, the NTA samples were soaked in the milli-Q water and calcinated in argon atmosphere at 400°C for 2 h using a tubular oven (Nabertherm, Germany).

#### *2.1.2 Mn$_x$O$_y$ electrodeposition*

The Ti/TiO$_2$ NTA-Mn$_x$O$_y$ electrodes were synthesized using anodic electrodeposition in a three-electrode setup at ambient temperature (i.e., 24±1°C), using stainless steel mesh as counter and Ag/AgCl (KCl 3M, Bioanalytical systems, the Netherlands) as reference electrode. 0.1 M MnSO$_4$ was selected as the precursor for the Mn$_x$O$_y$ electrodeposition. The solution also contained 0.05 or 0.5 M H$_2$SO$_4$ in order to determine the impact of the acid concentration on the Mn$_x$O$_y$ coating characteristics.

The electrodeposition was performed in the potentiostatic mode at 1.7 V/SHE (vs Standard Hydrogen Electrode). All the potentials reported in this work are expressed vs SHE and calculated according to Nernst's equation. To achieve equal thickness of the Mn$_x$O$_y$ coating in each deposition, the mass of the manganese oxide loading was estimated according to Faraday's law and the charge was limited to 13 C. To investigate the effect of the temperature treatment on the Mn$_x$O$_y$ coating, some Ti/TiO$_2$ NTA-Mn$_x$O$_y$ samples were also calcinated at 500°C for 1 h in a tubular oven (Nabertherm, Germany).

## *2.2 Ti/TiO$_2$ NTA – Mn$_x$O$_y$ material characterization*

The surface morphology of the synthesized materials was examined using an ultra-high-resolution field emission scanning electron microscopy (FESEM) (The Magellan 400L, FEI, US). The cross-section images were taken from the cracked layers after bending the samples. The crystal structure of the materials was determined by an X-ray powder diffractometer (X'Pert MPD, PANalytical, Netherlands) with Cu as Kα radiation source. The X-ray diffraction (XRD) patterns of the samples were recorded between 10 and 80° (2h) at a scan step size of 0.02°, and a time per step of 353 s. The chemical state analysis of the Mn$_x$O$_y$ coating was performed using X-ray photoelectron spectroscopy (XPS) using an X-ray photoelectron spectrometer (PHOIBOS 150, Specs, Germany). The average Mn valence state was determined by measuring the Mn 3s doublet peak separation as this method was reported to be more reliable compared to the analysis of Mn 2p peaks [42]. The values of Mn 3s doublet peak splitting for different average valence states of Mn reported in the literature include 4.5, 5.2, 5.4, and 5.8 eV for $MnO_2$ (Mn I), $Mn_2O_3$ (Mn III), $Mn_3O_4$ (Mn II, Mn III), and MnO (Mn II), respectively [42-44]. The detected O1s peaks were also deconvoluted and quantified to gain insight into the behavior of oxygen containing surface groups at different pH. The lowest energy peak (i.e., ~529.6 eV) is a typical response for oxygen in a transition metal oxide [45, 46]. The peak, observed at a higher binding energy (i.e., ~530.9 eV) is commonly attributed to the surface bound species such as oxygen in hydroxides [45, 46]. Finally, the peaks observed in the range between 532.5 and 535.5 eV typically represent weakly adsorbed oxygen species [46].

The electroactive surface area of the Ti/$TiO_2$ NTA-$Mn_xO_y$ electrodes was estimated by measuring the double layer capacitance ($C_{dl}$) observed during the cyclic voltammetry (CV) measurement in the 0.1 M $NaNO_3$. The scans were performed over a potential range of 0.4 – 1 V/SHE using scan rates between 30 and 2 mV s$^{-1}$. Values for $C_{dl}$ were determined by plotting the linear regression of the current versus the scan rate, according to the following equation:

$$\frac{I_a - I_c}{2} = C_{dl} \nu \qquad (eq.\ 1)$$

where $I_a$ and $I_c$ are the anodic and the cathodic currents observed in the forward and reversed scans (mA), respectively, and $\nu$ is the applied scan rate (mV s$^{-1}$). The electroactive surface area was then determined by dividing the capacitance by 60 mF cm$^{-2}$, which is considered as a standard value for the metal oxide based systems [47].

## 2.3 Sulfide removal tests

Electrochemical experiments were performed in an air-tight sealed non-divided electrochemical cell (100 mL). The synthesized Ti/$TiO_2$ NTA-$Mn_xO_y$ material was used as an anode, while Ag/AgCl (3M KCl) and Ti mesh (DeNora, Italy) served as the reference and counter electrodes, respectively. Electrochemical cells were sealed, and the headspace was under gentle nitrogen purge during the experiments to prevent the intrusion of oxygen and thus loss of sulfide to its oxidation to sulfate. The impact of potential on the sulfide removal was investigated by performing the experiments in the OC and at 0.4 V, 0.6 V or 0.8 V/SHE, using a deoxygenated 2.6 mM $NaNO_3$ as supporting electrolyte, with 2 mM $Na_2S$ at pH 12. The diluted supporting electrolyte solution (3.2 mS cm$^{-1}$) was purposely selected to simulate the conductivity of the real sewage (0.9-9 mS cm$^{-1}$) [48], as low ionic conductivity of real contaminated water is a limiting factor in electrochemical treatment systems. The impact of the initial sulfide concentration on the reaction kinetics was evaluated at 0.8 V/SHE by amending the supporting electrolyte with 0.9 mM, 2 mM or 3.2 mM of sulfide. All of the above-mentioned experiments were performed at the initial pH 12. To investigate the impact of pH, experiments were performed at the initial pH 8 in 2.6 mM $NaNO_3$ and 2 mM of $Na_2S$. As previously reported, lower initial pH of the solution has a minor effect on the reaction rate of sulfide oxidation with manganese oxides as it is typically followed by the rapid pH increase [49]. Therefore, determination of the pH impact on the oxidation rate required continuous addition of 0.1 M $HNO_3$ acid to the electrolyte. To investigate the performance of the Ti/TiO$_2$ NTA-Mn$_x$O$_y$ electrodes for sulfide removal in environmentally relevant conditions, experiments were carried out in real sewage that was filtered, deoxygenated and amended with 2 mM of sulfide.. All experiments were performed in duplicate, and values are expressed as mean with their standard errors.

The energy consumption was calculated as electric energy per order (Wh L$^{-1}$), required to reduce the concentration of sulfide by one order of magnitude in a unit volume of the treated solution according to the following equation:

$$E_{EO} = \frac{V \cdot I \cdot t}{v \cdot \log\left(\frac{C_0}{C}\right)} \qquad \text{(eq. 2)}$$

where V is cell voltage (V), I is measured current (A), v is the cell volume (L), $C_0$ and C are the initial and final concentration of sulfide, and t is the electrolysis time (h).

*2.4 Sample analysis*

The stability of the $Mn_xO_y$ coating was evaluated by measuring the concentration of the total dissolved manganese in both the $NaNO_3$ supporting electrolyte and sewage experiments. The measurement was performed with inductively coupled plasma-optical emission spectrometry (ICP-OES) (Agilent 5100, Agilent Technologies, US). The concentration of the dissolved sulfur species (i.e., $HS^-$, $S_2O_3^{2-}$, $SO_4^{2-}$) was measured by ion chromatography (IC, Dionex IC5000 (Dionex, USA). The experiments were performed in duplicates, the mean sulfide concentrations were normalized against the initial values and fitted to the first-order kinetics model with $R^2 \geq 0.99$. In order to compare sulfide oxidation kinetics observed using $Mn_xO_y$ based electrodes with different substrates (i.e., Ti/$TiO_2$ NTA and GF), the reaction rate was normalized to the geometric surface area.

The existing methods for $S^0$ measurement usually require a pre-concentration or extraction step with organic solvents [50, 51]. However, such step would be difficult to perform in this study because of different forms of the produced elemental sulfur (i.e., deposited at the electrode surface, and colloidal). Moreover, exposure of the Ti/$TiO_2$ NTA-$Mn_xO_y$ electrodes to organic solvents can affect catalytic properties and the subsequent performance of the coating. Thus, elemental sulfur was determined as the difference between the total sulfide added and the measured dissolved sulfur species. The ratio of the resulting elemental sulfur to sulfide removed during the experiment was considered as sulfur yield.

## Results and Discussion

### *3.1 Characterization of the Ti/TiO$_2$ NTA-Mn$_x$O$_y$ electrodes*

The anodization method employed resulted in uniform and well-aligned alignedTiO$_2$ NTAs, with outer average diameters of 80–100 nm, wall thickness of 7-10 nm and the average length of about 1 μm **(Figure S1).** The cross-section of the TiO$_2$ NTA substrate after the coating with Mn$_x$O$_y$ **(Figure 1a)** shows that anodic electrodeposition does not lead to any visible change in NTA morphology. The XRD patterns demonstrate that the NTA interlayer consists of pure tetragonal anatase phase, which is typically observed at the TiO$_2$ NTA after calcination at 400 °C **(Figure S2)** [52]. The diffraction peaks of the NTA-Mn$_x$O$_y$ have a similar position compared to the non-coated NTAs, which further confirms that the TiO$_2$ NTA structure was not affected by the electrodeposition process. Though demonstrating strong signal typical for Ti and TiO$_2$, the XRD diffraction pattern of the coated samples both with and without calcination showed no peaks characteristic for the Mn$_x$O$_y$ crystalline phase. The absence of the Mn$_x$O$_y$ signal in the XRD patterns was likely caused by overlapping of the substrate and the Mn$_x$O$_y$ peaks or by insufficient crystallinity of the Mn$_x$O$_y$ coating [53]. The presence of Mn$_x$O$_y$ was confirmed by the XPS analysis, as explained further in the text.

The anodic electrodeposition results in the penetration of Mn$_x$O$_y$ inside the NTAs **(Figure 1a)**, followed by the formation of a uniform Mn$_x$O$_y$ layer with the approximate thickness of 100 nm on the top of the NTAs. The top view of the Mn$_x$O$_y$ coatings synthesized at 0.05 M and 0.5 M H$_2$SO$_4$ concentration reveals visibly distinct Mn$_x$O$_y$ morphology. Lower acidity of the precursor solution yields smooth Mn$_x$O$_y$ coating **(Figure 1b)**, while high acid concentration leads to formation of the coating with rod-like morphology (Figure 1c). Shaker et al previously demonstrated that pH is the crucial parameter that can affect the structure, composition, and the morphology of the electrodeposited $Mn_xO_y$ coating [54].

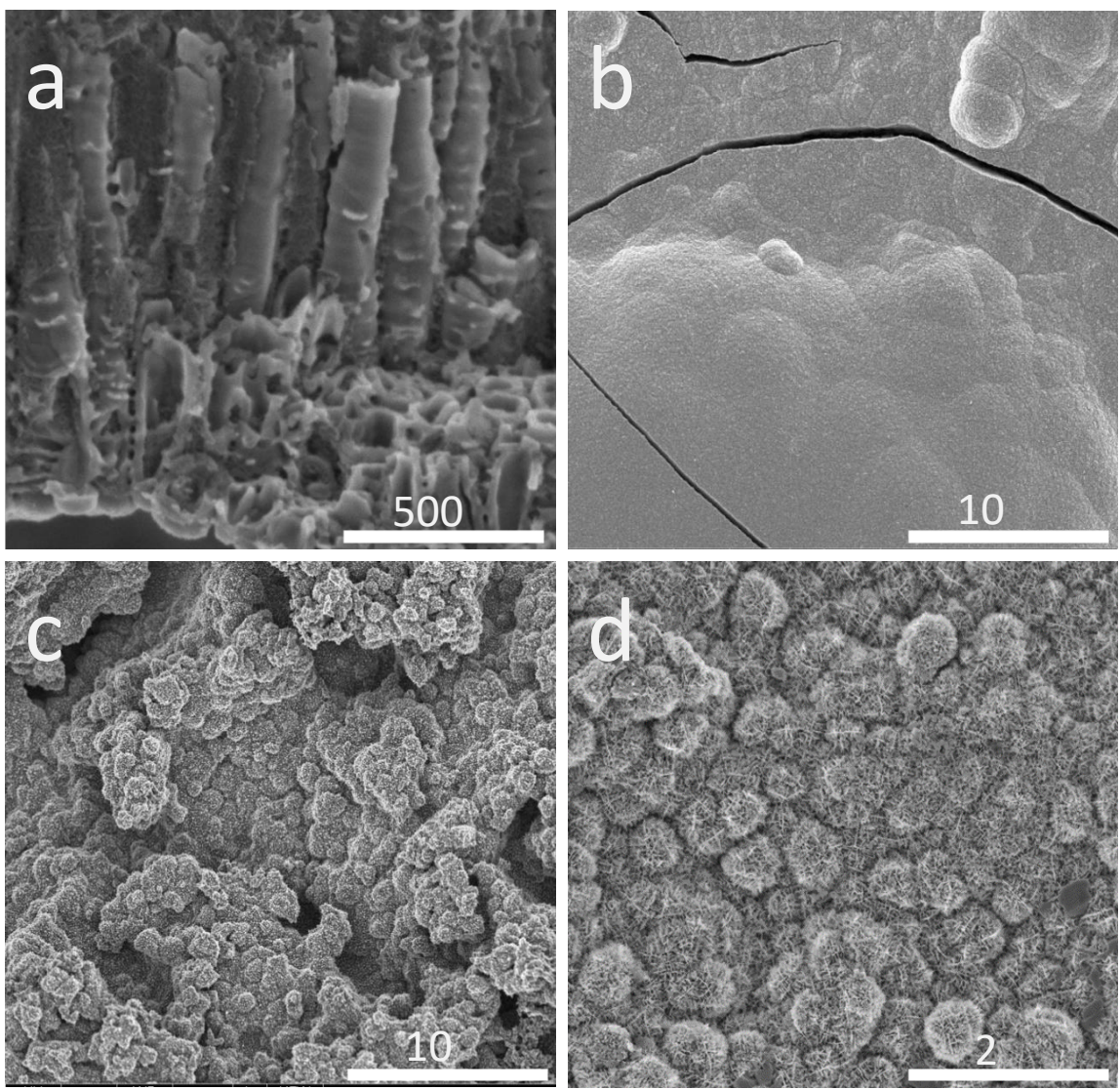

**Figure 1.** FESEM images of the cross section of self-organized $TiO_2$ nanotube array (NTA) filled with $Mn_xO_y$: **a)** top view of $TiO_2$ NTA coated with the $Mn_xO_y$ using precursor solution containing 0.1 M $MnSO_4$ and **b)** 0.05 M $H_2SO_4$, **c)** 0.5 M $H_2SO_4$, **d)** 0.5 M $H_2SO_4$ followed by the calcination at 500 ºC.

The difference in the pH of the deposition baths (e.g., pH 1.5 for 0.05 M $H_2SO_4$ and pH 0.5 for 0.5 M $H_2SO_4$) had a significant effect on the $Mn_xO_y$ coating morphology, as strongly acidic solution enhance the dissolution rate of the $Mn_xO_y$ and lead to the formation of a thinner coating with rod-like morphology [55]. The morphological difference of the two coatings may also result from the higher ionic conductivity of the strongly acidic bath (i.e., 179.3 mS cm$^{-1}$ for 0.5 M $H_2SO_4$, vs 24.5 mS cm$^{-1}$ for 0.05 M $H_2SO_4$), which leads to higher current during the anodic deposition of the $Mn_xO_y$, and thus significantly reduced deposition time (i.e., 42 s) compared to the less acidic bath (i.e., 74 s) **(Figure S3).** As was demonstrated in our previous study [38], prolonged electrodeposition leads to the smoother $Mn_xO_y$ coating surface as nanorods, formed upon initial nucleation, continue to grow in all directions and finally merge into adjacent growth centers. The smooth $Mn_xO_y$ coating is typically characterized by the low specific surface area due to the lack of relief in its structure. In addition to the electrodeposition bath acidity, subsequent calcination of the $Ti/TiO_2$ NTA/$Mn_xO_y$ had a significant impact on the material morphology, transforming the nanorods into needle like structures with more compact grains **(Figure 1d)**.

The composition of the $Mn_xO_y$ coating was investigated using the XPS analysis. The doublet peak splitting values reported in **Table 1** for the non-calcinated samples are typical for $MnO_2$ (Mn IV). This valence state is common for the $Mn_xO_y$ coatings synthesized via the anodic electrodeposition pathway in strong acidic media, that occurs through the following reaction [56]:

$$Mn^{2+}+2H_2O \rightarrow MnO_2+4H^{+}+2e^{-} \qquad (eq.\ 3)$$

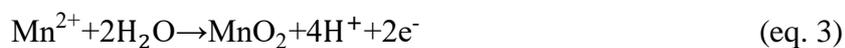

The doublet peak separation in the samples subjected to the calcination widened, which indicates the conversion of $MnO_2$ (Mn IV) into $Mn_2O_3$ (Mn III) **(Table 1, Figure S4)**. This phase transformation was reported to occur at 500ºC because of the desorption of the lattice oxygen and thermal decomposition according to the following pathway [57, 58]:

$$4MnO_2 \rightarrow 2Mn_2O_3 + O_2 \quad \text{(eq. 4)}$$

The impact of the NTA interlayer on the $Mn_xO_y$ coating morphology or composition could not be investigated because the $Mn_xO_y$ coating could not be deposited on a plane Ti plate in any of the investigated deposition conditions. Linear sweep voltammetry performed at the $TiO_2$ NTA indicates the clear peak at 1.7 V/SHE, which was attributed to the $Mn^{2+}$ oxidation to $MnO_2$, while bare Ti plate yields very low current with absence of any distinguishable peaks **(Figure S5)**. As can be seen from Figure S6, the resistance of the Ti plate is significantly higher compared to the anodized substrate with the $TiO_2$ NTAs. Therefore, limited charge transfer at the bare Ti substrate obstructed the nucleation and subsequent formation of the $Mn_xO_y$ coating, which was only possible on the highly conductive $TiO_2$ NTA interlayer.

### *3.2 Activity of the Ti/TiO$_2$ NTA-Mn$_x$O$_y$ electrodes towards sulfide oxidation*

All of the synthesized $Mn_xO_y$-coated electrodes, polarized at 0.8 V/SHE, caused a rapid sulfide decrease by 90 % within 3 h **(Figure 2)**. The main reaction product was elemental sulfur, with 94% yield of $S^0$. The produced elemental sulfur was deposited at the electrode surface, forming a loosely packed visible layer ($S_{8\,dep}$) **(Figure S7a)**. According to Luo et al., simultaneous participation of oxygen and manganese oxide in sulfide oxidation favors the formation of thiosulfate over elemental sulfur or sulfate [49]. Therefore, slight increase of $S_2O_3^{2-}$ concentrations (i.e., 0.1 mM or 6% of sulfide removed), observed in all performed experiments was a consequence of the presence of trace amounts of dissolved oxygen in the supporting electrolyte solution. Though several studies assumed that polysulfide is an important intermediate of sulfide oxidation with Mn oxides at basic pH, the solution stayed colorless throughout the experiment indicating the absence of polysulfide formation [59, 60].

The Ti/TiO$_2$ NTA-MnO$_2$ synthesized using lower acid concentration in the electrodeposition (i.e., 0.05 M H$_2$SO$_4$) performed slightly worse compared to the material synthesized in the presence of 0.5 M H$_2$SO$_4$ (i.e., 0.76±0.09 h$^{-1}$ and 0.98±0.1 h$^{-1}$, respectively) **(Figure 2)**. Smooth coating with little relief observed in the SEM images is typically associated with the lower active surface area and leads to a decreased catalytic activity [38]. Indeed, the electrochemically active surface area of the electrodes synthesized at higher acid concentration (i.e., 0.5 M H$_2$SO$_4$) was estimated at 408 cm$^2$, while lower acid concentration yielded material with slightly lower active surface are (i.e., 362 cm$^2$) **(Figure S8)**. Thus, the Ti/TiO$_2$ NTA – MnO$_2$ electrode synthesized using 0.5 M H$_2$SO$_4$ was selected for further experiments.

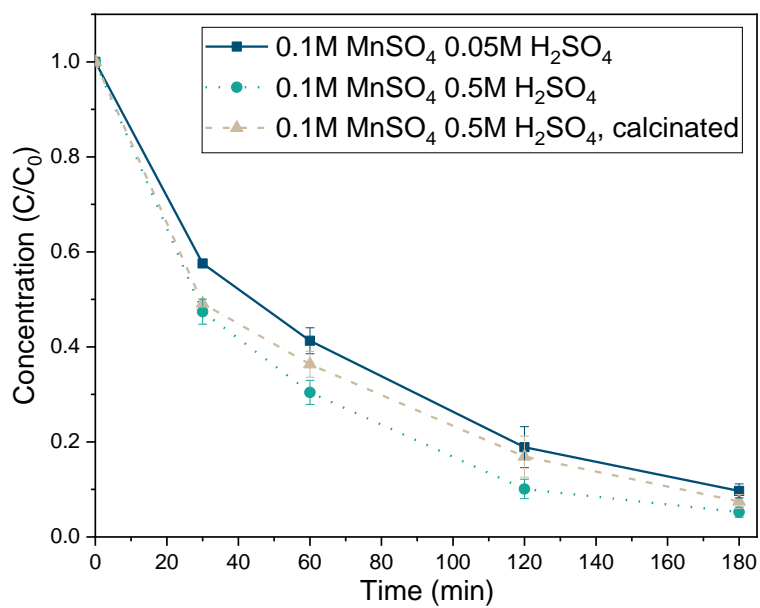

**Figure 2**. Decrease in the HS$^-$ concentration (C) normalized to the initial value (C$_0$) during the sulfide removal experiment at 0.8 V/SHE applied to the Ti/TiO$_2$ NTA-Mn$_x$O$_y$ electrodes.

Catalytic oxidation of sulfide is initiated with the adsorption of HS⁻ onto the $Mn_xO_y$ catalyst surface, followed by the formation of a surface complex, which is then oxidized to zero valent sulfur [60]:

$$Mn^{IV} + HS^- \rightleftharpoons Mn^{IV}S^- + H_2O \quad (eq.\ 5)$$

$$Mn^{IV}S^- \rightarrow Mn^{III}S \quad (eq.\ 6)$$

$$Mn^{III}S \rightarrow Mn^{II}S \quad (eq.\ 7)$$

$$Mn^{II}S \rightarrow Mn^{II} + S^0 \quad (eq.\ 8)$$

The electron transfer from Mn(IV) to sulfide occurs in two consecutive steps and involves the formation of the Mn(III) surface complex as a reaction intermediate. Being the intermediate, Mn(III) itself is capable of rapid catalytic oxidation of sulfide [61]. Although in the case of Mn(IV) the reaction mechanism requires one additional step, the transition between the surface complexes is so rapid that it does not significantly affect the removal rates, as evidenced from the results obtained with the non-calcined $MnO_2$ and calcined $Mn_2O_3$ coatings (i.e., 0.98±0.1 h⁻¹ and 0.83±0.01 h⁻¹ for Mn(IV) and Mn(III), respectively) **(Figure 2)**.

### *3.2.1 Impact of the applied anode potential*

The impact of the anode potential on sulfide oxidation was investigated by performing the experiments in the OC and at 0.4 V, 0.6, 0.8 V/SHE. As can be seen from Figure 3a, in the absence of any applied potential, the $Ti/TiO_2$ NTA–$MnO_2$ electrodes lead to a rapid drop in sulfide concentration by 40% within 30 min. Oxidation of sulfide occurs simultaneously with the reduction of the $MnO_2$ coating, which leads to the production of Mn(II) and its subsequent release into the electrolyte in the form of $Mn^{2+}$:

$$Mn^{II} \rightarrow Mn^{2+}_{aq} + \text{new surface site} \quad (eq.\ 9)$$

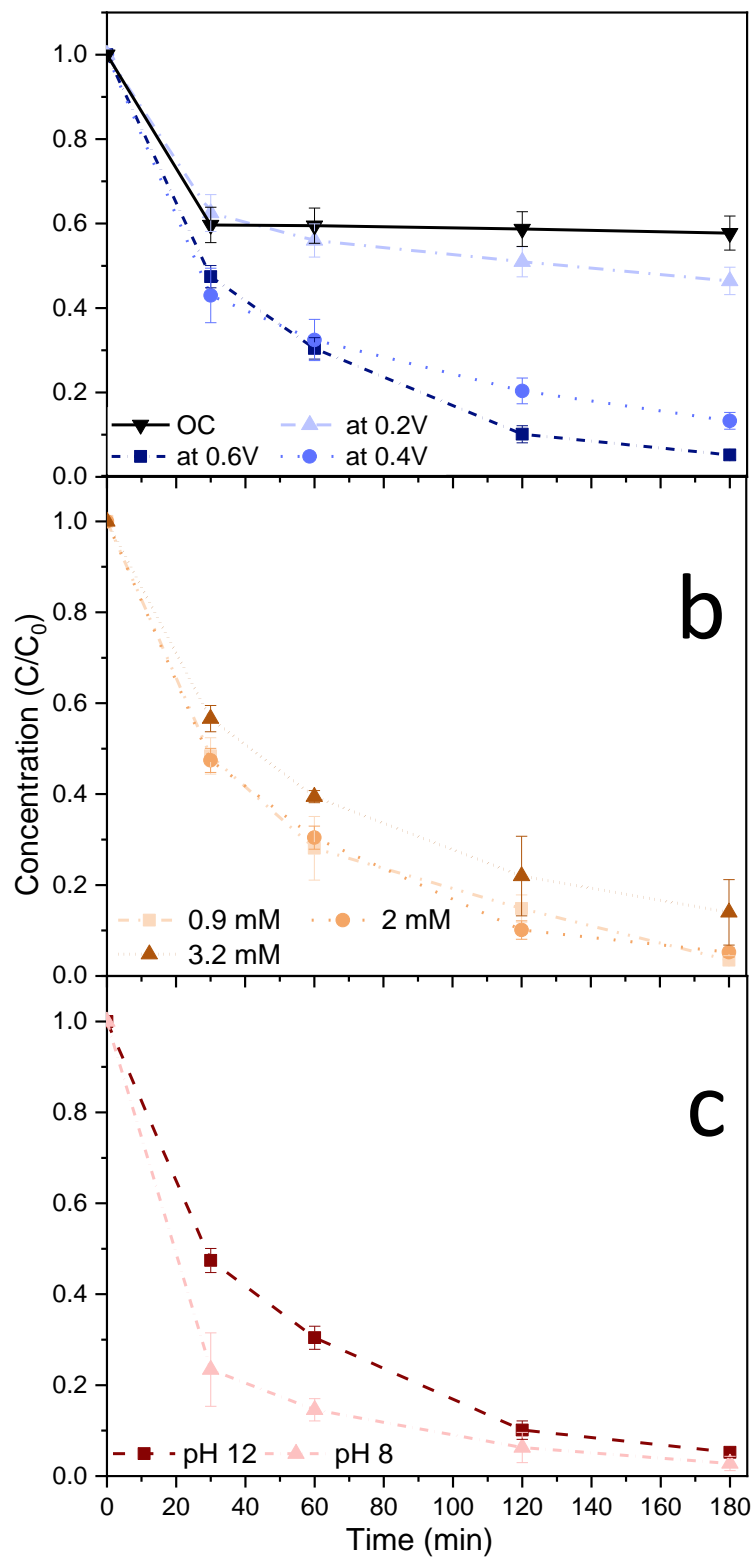

**Figure 3**. Decrease in the HS$^-$ concentration (C) normalized to the initial value (C$_0$) in the: **a)** open circuit (OC) or electrochemical sulfide removal experiments performed at various potentials, **b)** electrochemical experiments performed at 0.8 V/SHE, using various initial sulfide concentrations, **c)** electrochemical experiments performed at 0.8 V in 3 mM NaNO$_3$ supporting electrolyte amended with 2 mM of HS$^-$ at pH 8 and pH 12.

After this initial drop, sulfide concentration remained constant, indicating a complete depletion of the $MnO_2$ catalyst **(Figure 3a).** Complete dissolution of the coating was also confirmed by ICP-OES analysis of the supporting electrolyte, which detected substantial increase of total dissolved manganese concentration at the end of the experiment **(Table S1)**. The measured amount of the dissolved manganese (15.03 mg L$^{-1}$ in 100 mL, i.e., 1.503 mg) is slightly lower than the theoretical weight of the electrodeposited $MnO_2$ film, calculated according to the Faraday's law (i.e., 1.803 mg).

Polarization of the Ti/$TiO_2$ NTA-$MnO_2$ electrodes at the anodic potential as low as 0.4 V/SHE drastically minimized the loss of the $MnO_2$ catalyst. The amount of total dissolved manganese detected in the electrolyte (i.e., 0.07 mg) represented only 3.8% of the theoretical coating weight. The coating stabilization was achieved by recovering the reduced Mn(II) back to its initial oxidation state with the application of potential. The Mn 3s doublet peak separation in the samples after the sulfide removal experiments was equal to the pristine $TiO_2$ NTA-$MnO_2$ sample **(Table 2)**. However, such stabilization could not be achieved without an appropriate substrate such as $TiO_2$ NTA. The $TiO_2$ nanostructures not only improve the mechanical stability of the coating through better adhesion, but, more importantly, promote the charge transfer and enhance the conductivity of the electrode [30], which enables a rapid recovery of the Mn catalyst. Rapid restoration of the oxidation state of Mn maintained the catalytic activity of the Ti/$TiO_2$ NTA-$MnO_2$ electrode as was indicated by the gradual decrease of sulfide concentration during the entire experiment. Further increase of the potential significantly enhanced the sulfide oxidation kinetics (i.e., $0.21 \pm 0.07$ h$^{-1}$ at 0.4 V/SHE, $0.53 \pm 0.1$ h$^{-1}$ at 0.6 V/SHE and $0.98 \pm 0.2$ h$^{-1}$ at 0.8 V/SHE) **(Table S1)**. This was a consequence of a faster $MnO_2$ catalyst re-oxidation at higher potentials as the amount of dissolved Mn was decreased even further when potential was increased, with only 0.01 mg of Mn released at 0.8 V/SHE, which represents 0.53% of the theoretical coating weight **(Table S1)**. In addition, the reaction kinetics could also be accelerated though faster direct electrolysis of sulfide **(Figure 3a)**. For instance, elemental sulfur can be electrochemically oxidized to sulfate at standard redox potential of 0.357 V/SHE, however, no increase of the dissolved sulfur species could be detected [20].

### *3.2.2 Impact of the initial sulfide concentration and pH*

The initial sulfide concentration had a pronounced effect on the sulfide oxidation rate **(Figure 3b).** Lowering the initial sulfide concentration from 2 mM to 0.9 mM did not affect the reaction kinetics, as first-order rate constants were the same (i.e., $1.04 \pm 0.05$ $h^{-1}$ and $0.98 \pm 0.2$ $h^{-1}$, respectively). Increase in the initial sulfide concentration to 3.9 mM led to slower sulfide oxidation kinetics, with rate constants of $0.62 \pm 0.1$ $h^{-1}$ **(Figure 3b, Table S2)**. Slower removal rates at higher $HS^-$ concentration confirms that the reaction occurs through the formation of an inner sphere complex with $MnO_2$. The inner sphere mechanism is not limited by the mass transfer, instead the rate of the reaction is determined by the rate at which $HS^-$ ions can form complexes with the $MnO_2$ catalyst [62]. As can be seen from Table S2, higher $HS^-$ concentration had no negative impact on the $MnO_2$ coating stability as the total dissolved manganese detected in the electrolyte (i.e., 0.01 mg) represented less than 0.6% of the theoretical weight of the deposited $MnO_2$ coating.

At pH 8, sulfide was removed at a higher rate ($1.32 \pm 0.2$ $h^{-1}$) compared with the pH 12 experiments ($0.98 \pm 0.2$ $h^{-1}$) **(Figure 3c, Table S3)**. The pH-dependence of the sulfide oxidation rate further supports an inner-sphere complex formation between $HS^-$ and the $MnO_2$ coating [60].

The catalytic activity and stability of the MnO$_2$ catalysts is generally very sensitive towards the pH of the electrolyte [63]. The surface groups of manganese oxides are amphoteric, meaning they can function both as an acid and a base [64]. Hence, the catalyst surface can undergo protonation and/or deprotonation depending on the value of the pH of the solution [65]:

$$Mn^{IV}OH \leftrightarrow Mn^{IV}O^- + H^+ \quad pK_1 = 8.2 \quad (eq.\ 10)$$

On the other hand, sulfide speciation is also dependent on the pH as can be seen from the sulfide ionization equilibrium [7]:

$$H_2S_{(aq)} \leftrightarrow HS^- \quad pK_1 = 6.88 \quad (eq.\ 11)$$

The formation of a complex is affected by the relative distribution of different sites on the manganese oxide surface and of sulfide in the solution [49, 60]. According to Zhu et al., the pH range between 7 and 8 enables the formation [Mn$^{IV}$OH][HS$^-$] complex, which leads to faster sulfide oxidation kinetics compared to [Mn$^{IV}$OH][H$_2$S] and [Mn$^{IV}$O$^-$][HS$^-$] [49]. The increase of hydroxide containing surface groups in samples exposed to electrolyte at basic pH was also confirmed by the XPS analysis (**Table 2, Figure S9**). Even though the final product of the treatment at both pH 8 and 12 was elemental sulfur, at pH 12 sulfur remained adsorbed at the anode surface (S$_{8\ dep}$), while at pH 8 sulfur was formed and instantly desorbed from the anode surface and released into the electrolyte producing colloidal sulfur particles of ~ 0.2 μm (S$_{8\ col}$) (**Figure S10**). Adsorption of the elemental sulfur to the Ti/TiO$_2$NTA-MnO$_2$ electrode surface at pH 12 was confirmed by the recorded SEM and EDX **(Figure 4a)**, as well as XPS analysis **(Figure S11)**. The SEM images of the electrode surface after the sulfide removal test at pH 8 demonstrate that the MnO$_2$ morphology remained unchanged **(Figure 4b)**. Moreover, neither EDX nor XPS detected the presence of sulfur, thus further evidencing that when controlling the supporting electrolyte pH at pH 8, the formed elemental sulfur was not electrodeposited but released as $S_{8\ col}$. The difference in the forms of produced sulfur indicates the shift in the $S_8$ formation mechanism. As was mentioned in Equation 8, further oxidation of surface complexes of sulfide ion and manganese yields zero valent sulfur, $S^0$. At pH 8, zero valent sulfur is desorbed from the anode surface and released into the solution, where it undergoes further complexation to $S_{8\ col}$ [60]. On the contrary, at pH 12, formation of elemental sulfur ($S_{8\ dep}$) becomes diffusion controlled and occurs at the electrode surface.

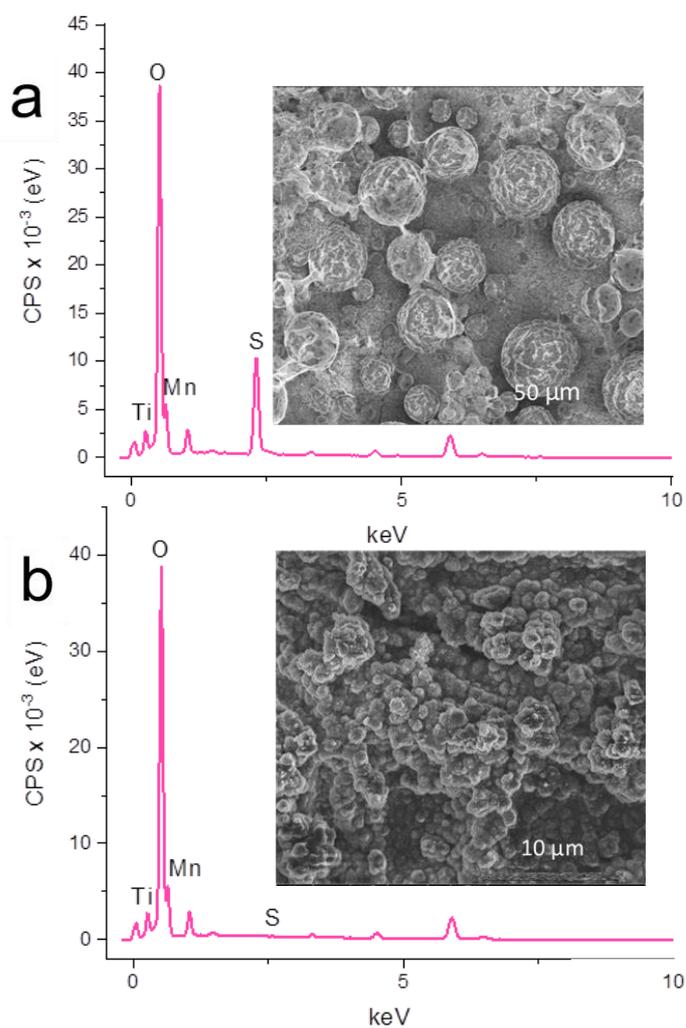

**Figure 4**. FESEM images and EDX spectra of the: **a)** Ti/TiO$_2$ NTA-MnO$_2$ electrode after the electrochemical sulfide removal test performed at pH 12, **b)** Ti/TiO$_2$ NTA-MnO$_2$ electrode after the electrochemical sulfide removal test performed at pH 8.

Direct electrooxidation of sulfide to elemental sulfur is typically associated with the gradual loss of performance, which occurs as a result electrode passivation with the isolating layer of elemental sulfur [15, 16, 20]. In electrooxidation at pH 12 in which $S_{8\ dep}$ remained at the anode surface, rapid decrease in sulfide removal rates was observed within three consecutive applications, i.e., 0.81±0.03 h$^{-1}$, 0.41±0.06 h$^{-1}$, and 0.19±0.08 h$^{-1}$ in the first, second and third cycle **(Figure 5, Table S4)**. Moreover, the electrode passivation was further supported by the increase in the energy consumption in each subsequent cycle (i.e., 0.09 Wh L$^{-1}$, 0.18 Wh L$^{-1}$ and 0.79 Wh L$^{-1}$ in the first, second and third application, respectively) **(Table S4)**. Similar passivation with elemental sulfur was noted in the repetitive applications of the Ti/TiO$_2$NTA-MnO$_2$ anode in the experiments without pH control, as pH was rapidly increased from the initial pH 8 to pH 12 (data not shown). On the contrary, when the pH was controlled at pH 8, the Ti/TiO$_2$NTA-MnO$_2$ electrode demonstrated rapid and robust sulfide removal in the consecutive cycles, with unchanged sulfide removal rate constants (i.e., 1.01±0.04 h$^{-1}$, 0.96±0.06 h$^{-1}$, and 1.06±0.1 h$^{-1}$ in the first, second and third application) and energy consumption (i.e., 0.05 Wh L$^{-1}$, 0.044 Wh L$^{-1}$ and 0.042 Wh L$^{-1}$ in the first, second and third application, respectively), demonstrating that the electrode passivation was effectively avoided **(Figure 5, Table S4)**.

### *3.3 Sulfide oxidation at the Ti/TiO$_2$ NTA-Mn$_x$O$_y$ electrodes in real sewage*

Sulfide removal kinetics in the experiments with real sewage was slightly slower (0.69±0.06 h$^{-1}$) compared to the tests carried out in the NaNO$_3$ supporting electrolyte (1.01±0.04 h$^{-1}$) **(Figure 5, Table S4)**, while the energy consumption of the system increased (i.e., 0.05 Wh L$^{-1}$ and 0.29 Wh L$^{-1}$ in the supporting NaNO$_3$ electrolyte and in real sewage, respectively). This can be explained by the participation of the MnO$_2$ coating in reactions other than sulfide oxidation. A slight decrease (i.e., 8% removal) in chemical oxygen demand (COD) of was observed at the end of the experiments, indicating a partial oxidation of the organic matter by the $MnO_2$ coating. Nevertheless, sulfide was still completely removed within 3 h, indicating high selectivity of $TiO_2$ NTA-$MnO_2$ electrode even in the case of complex matrix such as real sewage. Moreover, phosphate, typically present in wastewater, can block the $MnO_2$ surface sites available for the reaction with sulfide and inhibit sulfide removal [66]. The competition between sulfide and phosphate for the $MnO_2$ sites were further confirmed by the slight decrease of phosphate concentration (i.e., from 0.08 mM to 0.04 mM). Notwithstanding the presence of organics and inorganics in real sewage that are known to react with the Mn-oxide [67, 68] and can cause reductive dissolution of the $MnO_2$ coating, the electrodes remained stable and no release of $Mn^{2+}$ ions was detected in the ICP-OES analyses.

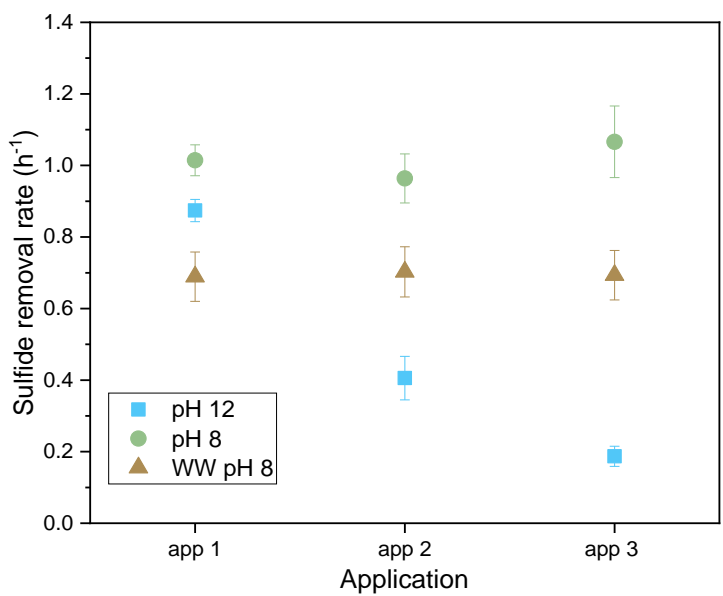

**Figure 5**. Observed first-order sulfide removal rates ($h^{-1}$) at the Ti/$TiO_2$ NTA-$Mn_xO_y$ electrodes applied over three subsequent cycles in synthetic electrolyte containing 3 mM $NaNO_3$ and 2 mM of $HS^-$, at pH 12 and pH 8, and in the real sewage at pH 8.

Given that the buffering capacity of the real sewage avoided an increase in pH during the experiment and maintained it at pH 8, the formation of elemental sulfur proceeded in the same manner as in the experiments performed using $NaNO_3$ electrolyte at pH 8. The final product of the sulfide oxidation in real swage was zero valent sulfur, that underwent complexation to $S_{8\ col}$ in the bulk of the electrolyte. Thus, the passivation of the anode surface with elemental sulfur was completely avoided, and there was no decrease in the sulfide removal rates in subsequent applications (i.e., 0.69±0.06 $h^{-1}$, 0.70±0.065 $h^{-1}$, 0.69±0.09 $h^{-1}$ in the first, second and third cycle) and the energy consumption of the system remained constant (i.e., 0.29 Wh $L^{-1}$, 0.26 Wh $L^{-1}$, 0.31 Wh $L^{-1}$ in the first, second and third cycle) **(Figure 5, Table S4).**

The deposition of sulfur at the electrode surface might be considered beneficial for application in sewer systems as it enables complete separation of sulfur from the stream, thus, avoiding the problems with sulfide reformation [13]. Moreover, direct oxidation of sulfide to sulfur requires low energy input, especially compared to the indirect sulfide oxidation with electrochemically produced oxygen, or via direct electron transfer to sulfate [69]. On the other hand, several studies mentioned that the efficiency of direct sulfide oxidation is significantly deteriorated in sewage with low sulfur content due to mass transfer limitation, which is a major issue considering that average sulfide concentration is sewage rarely exceeds 3 mM [18]. Besides, gradual loss of performance associated with the sulfur deposition is another important drawback which prevents the implementation of this approach in practice [15, 16, 20]. Previously proposed strategies aimed at the mitigation of the electrode passivation such as extraction of sulfur with organic solvents [70], use of surfactants [71], and cathodic dissolution of the electrodeposited sulfur [15, 16] are either incapable of full restoration of sulfide removal capacity, economically unattractive or environmentally hazardous. Hence, switching to electrochemical production of colloidal sulfur at low applied potentials can help maintain low energy requirements of direct sulfide oxidation to sulfur, while avoiding problems with passivation and decreased efficiency in waste streams with low sulfide content. Even though colloidal sulfur remains in the stream, it is unlikely to lead to sulfide reformation due to its low bioavailability, especially when compared to other dissolved sulfur species [72].

## 4. Conclusion

The $TiO_2$ NTA interlayer enabled the electrodeposition of the $Mn_xO_y$ coating by significantly increasing the conductivity and specific surface area of the bare Ti plate. The electrodeposition of the $Mn_xO_y$ led to filling of the $TiO_2$ nanotubes without damaging their morphology. The $Ti/TiO_2NTA-MnO_2$ anode demonstrated excellent catalytic activity towards sulfide oxidation, yielding a thirty-fold higher normalized reaction rate compared to $Mn_2O_3$-coated GF anodes (i.e., 175 x $10^4$ $m^3$ $h^{-1}$ $m^{-2}$ and 6 x $10^4$ $m^3$ $h^{-1}$ $m^{-2}$ for $Ti/TiO_2NTA-MnO_2$ and $GF-Mn_2O_3$, respectively) [38]. Oxidation of sulfide on $MnO_2$ coating occurred through formation of the inner sphere complex, yielding elemental sulfur as the final product. Coating dissolution, observed in the OC experiment due to the $Mn_xO_y$ reduction, was effectively prevented by the continuous recovery of the catalyst under low applied potentials. The pH of the electrolyte not only affected the reaction kinetics, but also determined the form of the resulting elemental sulfur. At pH 12, zero valent sulfur remained adsorbed at the anode surface, whereas at pH 8 it was desorbed and further complexed to $S_8$, thus avoiding completely the anode passivation. The electrochemical production of colloidal sulfur instead of its electrodeposition at the anode surface enabled a sustained (electro)catalytic activity of the $Ti/TiO_2NTA-MnO_2$ anode in the subsequent applications in $NaNO_3$ electrolyte at constant pH 8, as well as in the real sewage in which constant pH 8 was maintained by the buffering capacity of sewage.

This unprecedented performance of the Ti/TiO$_2$NTA-Mn$_x$O$_y$ anodes developed in this study enabled rapid sulfide oxidation and selectivity towards colloidal sulfur in the conditions relevant for sewage treatment and other processes where removal of dissolved sulfide is of interest (e.g., sulfide removal from alkaline solutions used in biogas scrubbers, at pH 8-9). Continuous electrocatalytic recovery of the coating is of crucial importance for reusing a catalyst that would be rapidly depleted in a conventional homogeneous catalytic process. Electrocatalytic sulfide removal using Ti/TiO$_2$NTA-Mn$_x$O$_y$ anodes overcomes two major limitations of conventional direct electrochemical sulfide oxidation - low efficiency for diluted sulfide solutions, and electrode passivation with sulfur, while maintaining low energy input requirements (0.29 Wh L$^{-1}$). Taking into consideration the easiness of the electrode synthesis procedure, low cost of the earth-abundant manganese, low energy consumption, low bioavailability of electrochemically produced sulfur, modularity and possibility of automated, remote operation of the electrooxidation of sulfide at Ti/TiO$_2$NTA-MnO$_2$ electrodes, the proposed approach can potentially become a game-changing technology of sulfide control in sewer systems and other applications.

## Acknowledgments

The authors would like to acknowledge ERC Starting Grant project ELECTRON4WATER (Three-dimensional nanoelectrochemical systems based on low-cost reduced graphene oxide: the next generation of water treatment systems), project number 714177. ICRA researchers thank funding from CERCA program.## References

[1] N.N. Greenwood, A. Earnshaw, 15 - Sulfur, Chemistry of the Elements (Second Edition), Butterworth-Heinemann, Oxford, 1997, pp. 645-746. https://doi.org/10.1016/B978-0-7506-3365-9.50021-3

[2] C.A. Bondi, J.L. Marks, L.B. Wroblewski, H.S. Raatikainen, S.R. Lenox, K.E. Gebhardt, Human and Environmental Toxicity of Sodium Lauryl Sulfate (SLS): Evidence for Safe Use in Household Cleaning Products, Environ Health Insights, 9 (2015) 27-32 10.4137/EHI.S31765.

[3] W. Rauch, M. Kleidorfer, Replace contamination, not the pipes, Science, 345 (2014) 734-735 10.1126/science.1257988.

[4] A. Sarti, M. Zaiat, Anaerobic treatment of sulfate-rich wastewater in an anaerobic sequential batch reactor (AnSBR) using butanol as the carbon source, Journal of Environmental Management, 92 (2011) 1537-1541 https://doi.org/10.1016/j.jenvman.2011.01.009.

[5] J. Vollertsen, T. Hvitved-Jacobsen, I. McGregor, R. Ashley, Aerobic microbial transformations of pipe and silt trap sediments from combined sewers, Water Science and Technology, 38 (1998) 249-256 https://doi.org/10.1016/S0273-1223(98)00756-2.

[6] L.L. Barton, G.D. Fauque, Chapter 2 Biochemistry, Physiology and Biotechnology of Sulfate-Reducing Bacteria, Advances in Applied Microbiology, Academic Press2009, pp. 41-98. https://doi.org/10.1016/S0065-2164(09)01202-7

[7] M. Abdollahi, A. Hosseini, Hydrogen Sulfide, in: P. Wexler (Ed.) Encyclopedia of Toxicology (Third Edition), Academic Press, Oxford, 2014, pp. 971-974. https://doi.org/10.1016/B978-0-12-386454-3.00513-3

[8] B. Doujaiji, J.A. Al-Tawfiq, Hydrogen sulfide exposure in an adult male, Ann Saudi Med, 30 (2010) 76-80 10.4103/0256-4947.59379.27

[9] Å.D. Austigard, K. Svendsen, K.K. Heldal, Hydrogen sulphide exposure in waste water treatment, J Occup Med Toxicol, 13 (2018) 10-10 10.1186/s12995-018-0191-z.

[10] A.K. Parande, P.L. Ramsamy, S. Ethirajan, C.R.K. Rao, N. Palanisamy, Deterioration of reinforced concrete in sewer environments, Proceedings of the Institution of Civil Engineers - Municipal Engineer, 159 (2006) 11-20 10.1680/muen.2006.159.1.11.

[11] J. Monteny, E. Vincke, A. Beeldens, N. De Belie, L. Taerwe, D. Van Gemert, W. Verstraete, Chemical, microbiological, and in situ test methods for biogenic sulfuric acid corrosion of concrete, Cement and Concrete Research, 30 (2000) 623-634 https://doi.org/10.1016/S0008-8846(00)00219-2.

[12] A. Romanova, D.M. Mahmoodian, M. Alani, Influence and Interaction of Temperature, H2S and pH on Concrete Sewer Pipe Corrosion, 2014.

[13] I. Pikaar, E.M. Likosova, S. Freguia, J. Keller, K. Rabaey, Z. Yuan, Electrochemical Abatement of Hydrogen Sulfide from Waste Streams, Critical Reviews in Environmental Science and Technology, 45 (2015) 1555-1578 10.1080/10643389.2014.966419.

[14] D. Firer, E. Friedler, O. Lahav, Control of sulfide in sewer systems by dosage of iron salts: Comparison between theoretical and experimental results, and practical implications, Science of The Total Environment, 392 (2008) 145-156 https://doi.org/10.1016/j.scitotenv.2007.11.008.

[15] P.K. Dutta, R.A. Rozendal, Z. Yuan, K. Rabaey, J. Keller, Electrochemical regeneration of sulfur loaded electrodes, Electrochemistry Communications, 11 (2009) 1437-1440 https://doi.org/10.1016/j.elecom.2009.05.024.

[16] N. Sergienko, E. Irtem, O. Gutierrez, J. Radjenovic, Electrochemical removal of sulfide on porous carbon-based flow-through electrodes, Journal of Hazardous Materials, 375 (2019) 19-25 https://doi.org/10.1016/j.jhazmat.2019.04.033.

[17] J. Radjenovic, D.L. Sedlak, Challenges and Opportunities for Electrochemical Processes as Next-Generation Technologies for the Treatment of Contaminated Water, Environmental Science & Technology, 49 (2015) 11292-11302 10.1021/acs.est.5b02414.

[18] I. Pikaar, R.A. Rozendal, Z. Yuan, J. Keller, K. Rabaey, Electrochemical sulfide removal from synthetic and real domestic wastewater at high current densities, Water Research, 45 (2011) 2281-2289 https://doi.org/10.1016/j.watres.2010.12.025.

[19] I. Pikaar, R.A. Rozendal, Z. Yuan, J. Keller, K. Rabaey, Electrochemical sulfide oxidation from domestic wastewater using mixed metal-coated titanium electrodes, Water Research, 45 (2011) 5381-5388 https://doi.org/10.1016/j.watres.2011.07.033.

[20] B. Ateya, F. AlKharafi, A. Al-Azab, Electrodeposition of Sulfur from Sulfide Contaminated Brines, Electrochemical and Solid State Letters - ELECTROCHEM SOLID STATE LETT, 6 (2003)  10.1149/1.1599686.

[21] B.G. Ateya, F.M. Al-Kharafi, Anodic Oxidation of Sulfide Ions from Chloride Brines, Electrochemistry Communications, 4 (2002) 231-238 10.1016/S1388-2481(02)00254-0.

[22] K. Wlodarchak, D. Skutt, G. Deshinsky, A. Chan, E. Pedersen, Odor Reduction in a Wastewater Treatment Plant Using Potassium Permanganate, Proceedings of the Water Environment Federation, 2002 (2002) 72-81 10.2175/193864702785139999.

[23] F. Cadena, R.W. Peters, Evaluation of Chemical Oxidizers for Hydrogen Sulfide Control, Journal (Water Pollution Control Federation), 60 (1988) 1259-1263

[24] B. Zhang, L. Sun, Why nature chose the Mn4CaO5 cluster as water-splitting catalyst in photosystem II: a new hypothesis for the mechanism of O-O bond formation, Dalton Trans, 47 (2018) 14381-14387 10.1039/c8dt01931b.

[25] G. Davies, Some aspects of the chemistry of manganese(III) in aqueous solution, Coordination Chemistry Reviews, 4 (1969) 199-224 https://doi.org/10.1016/S0010-8545(00)80086-7.

[26] I. Roger, M.A. Shipman, M.D. Symes, Earth-abundant catalysts for electrochemical and photoelectrochemical water splitting, Nature Reviews Chemistry, 1 (2017) 0003 10.1038/s41570-016-0003.

[27] W. Xiao, P. Zhou, X. Mao, D. Wang, Ultrahigh aniline-removal capacity of hierarchically structured layered manganese oxides: trapping aniline between interlayers, Journal of Materials Chemistry A, 3 (2015) 8676-8682 10.1039/C5TA01305D.

[28] J. Wang, J. Li, C. Jiang, P. Zhou, P. Zhang, J. Yu, The effect of manganese vacancy in birnessite-type MnO2 on room-temperature oxidation of formaldehyde in air, Applied Catalysis B: Environmental, 204 (2017) 147-155 https://doi.org/10.1016/j.apcatb.2016.11.036.

[29] J. Wang, G. Zhang, P. Zhang, Graphene-assisted photothermal effect on promoting catalytic activity of layered MnO2 for gaseous formaldehyde oxidation, Applied Catalysis B: Environmental, 239 (2018) 77-85 https://doi.org/10.1016/j.apcatb.2018.08.008.

[30] A. Massa, S. Hernández, A. Lamberti, C. Galletti, N. Russo, D. Fino, Electro-oxidation of phenol over electrodeposited MnOx nanostructures and the role of a TiO2 nanotubes interlayer, Applied Catalysis B: Environmental, 203 (2017) 270-281 https://doi.org/10.1016/j.apcatb.2016.10.025.

[31] Y.-q. Wang, B. Gu, W.-l. Xu, Electro-catalytic degradation of phenol on several metal-oxide anodes, Journal of Hazardous Materials, 162 (2009) 1159-1164 https://doi.org/10.1016/j.jhazmat.2008.05.164.

[32] M. Nakayama, M. Shamoto, A. Kamimura, Surfactant-Induced Electrodeposition of Layered Manganese Oxide with Large Interlayer Space for Catalytic Oxidation of Phenol, Chemistry of Materials, 22 (2010) 5887-5894 10.1021/cm101970b.

[33] V.P. Santos, M.F.R. Pereira, J.J.M. Órfão, J.L. Figueiredo, The role of lattice oxygen on the activity of manganese oxides towards the oxidation of volatile organic compounds, Applied Catalysis B: Environmental, 99 (2010) 353-363 https://doi.org/10.1016/j.apcatb.2010.07.007.

[34] S.S.T. Bastos, J.J.M. Órfão, M.M.A. Freitas, M.F.R. Pereira, J.L. Figueiredo, Manganese oxide catalysts synthesized by exotemplating for the total oxidation of ethanol, Applied Catalysis B: Environmental, 93 (2009) 30-37 https://doi.org/10.1016/j.apcatb.2009.09.009.

[35] I. Devai, R.D. Delaune, Effectiveness of selected chemicals for controlling emission of malodorous sulfur gases in sewage sludge, Environmental technology, 23 (2002) 319-329 10.1080/09593332508618412.

[36] D. Thomas, Reducing Hydrogen Sulfide (H2S) Concentrations at Wastewater Collection Systems and Treatment Facilities using Chemical Oxidation, (2007)

[37] S. Asaoka, H. Okamura, Y. Akita, K. Nakano, K. Nakamoto, K. Hino, T. Saito, S. Hayakawa, M. Katayama, Y. Inada, Regeneration of manganese oxide as adsorption sites for hydrogen sulfide on granulated coal ash, Chemical Engineering Journal, 254 (2014) 531-537 https://doi.org/10.1016/j.cej.2014.06.005.

[38] N. Sergienko, J. Radjenovic, Manganese oxide-based porous electrodes for rapid and selective (electro)catalytic removal and recovery of sulfide from wastewater, Applied Catalysis B: Environmental, 267 (2020) 118608 https://doi.org/10.1016/j.apcatb.2020.118608.

[39] E. Petrucci, D. Montanaro, M. Orsini, G. Sotgiu, Micro- and nanostructured TiO2 substrate: Influence on the electrocatalytic properties of manganese oxide based electrodes, Journal of Electroanalytical Chemistry, 808 (2018) 380-386 https://doi.org/10.1016/j.jelechem.2017.07.008.

[40] G. Zhao, X. Cui, M. Liu, P. Li, Y. Zhang, T. Cao, H. Li, Y. Lei, L. Liu, D. Li, Electrochemical Degradation of Refractory Pollutant Using a Novel Microstructured TiO2 Nanotubes/Sb-Doped SnO2 Electrode, Environmental Science & Technology, 43 (2009) 1480-1486 10.1021/es802155p.

[41] H. An, H. Cui, W. Zhang, J. Zhai, Y. Qian, X. Xie, Q. Li, Fabrication and electrochemical treatment application of a microstructured TiO2-NTs/Sb–SnO2/PbO2 anode in the degradation of C.I. Reactive Blue 194 (RB 194), Chemical Engineering Journal, 209 (2012) 86-93 https://doi.org/10.1016/j.cej.2012.07.089.

[42] P. Hosseini-Benhangi, C.H. Kung, A. Alfantazi, E. Gyenge, Controlling the Interfacial Environment in the Electrosynthesis of MnOx Nanostructures for High Performance Oxygen Reduction/Evolution Electrocatalysis, ACS Applied Materials & Interfaces, 9 (2017) https://doi.org/10.1021/acsami.7b05501.

[43] M. Chigane, M. Ishikawa, Manganese Oxide Thin Film Preparation by Potentiostatic Electrolyses and Electrochromism, Journal of The Electrochemical Society - J ELECTROCHEM SOC, 147 (2000) https://doi.org/10.1149/1.1393515.

[44] L. Zhong Zhao, V. Young, XPS studies of carbon supported films formed by the resistive deposition of manganese, Journal of Electron Spectroscopy and Related Phenomena, 34 (1984) 45-54 https://doi.org/10.1016/0368-2048(84)80058-4.

[45] A.J. Naylor, E. Makkos, J. Maibach, N. Guerrini, A. Sobkowiak, E. Björklund, J.G. Lozano, A.S. Menon, R. Younesi, M.R. Roberts, K. Edström, M.S. Islam, P.G. Bruce, Depth-dependent oxygen redox activity in lithium-rich layered oxide cathodes, Journal of Materials Chemistry A, 7 (2019) 25355-25368 10.1039/C9TA09019C.

[46] J. Dupin, D. Gonbeau, P. Vinatier, A. Levasseur, Systematic XPS studies of metal oxides, hydroxides and peroxides, Physical Chemistry Chemical Physics, 2 (2000) 1319-1324

[47] S. Nayak, B.P. Chaplin, Fabrication and characterization of porous, conductive, monolithic $Ti_4O_7$ electrodes, Electrochimica Acta, 263 (2018) 299-310 https://doi.org/10.1016/j.electacta.2018.01.034.

[48] VanLondon, Conductivity Guide, 2018.

[49] Y. Luo, S. Li, W. Tan, G. Qiu, Oxidation and Catalytic Oxidation of Dissolved Sulfide by Manganite in Aqueous Systems, Clays and Clay Minerals, 65 (2017) 299-309 10.1346/CCMN.2017.064066.

[50] A. Kamyshny Jr, C.G. Borkenstein, T.G. Ferdelman, Protocol for Quantitative Detection of Elemental Sulfur and Polysulfide Zero-Valent Sulfur Distribution in Natural Aquatic Samples, Geostandards and Geoanalytical Research, 33 (2009) 415-435 https://doi.org/10.1111/j.1751-908X.2009.00907.x.

[51] Y.-W. Chen, H.A. Joly, N. Belzile, Determination of elemental sulfur in environmental samples by gas chromatography-mass spectrometry, Chemical Geology, 137 (1997) 195-200 https://doi.org/10.1016/S0009-2541(96)00163-5.

[52] P. Roy, S. Berger, P. Schmuki, $TiO_2$ Nanotubes: Synthesis and Applications, Angewandte Chemie International Edition, 50 (2011) 2904-2939 https://doi.org/10.1002/anie.201001374.

[53] C. Holder, R. Schaak, Tutorial on Powder X-ray Diffraction for Characterizing Nanoscale Materials, ACS Nano, 13 (2019) 7359-7365 10.1021/acsnano.9b05157.

[54] M.N. Shaker, S.A. Bonke, J. Xiao, M.A. Khan, R.K. Hocking, M.F. Tesch, Insight into pH-Dependent Formation of Manganese Oxide Phases in Electrodeposited Catalytic Films Probed by Soft X-Ray Absorption Spectroscopy, ChemPlusChem, 83 (2018) 721-727 10.1002/cplu.201800055.

[55] E.B. Godunov, A.D. Izotov, I.G. Gorichev, Dissolution of Manganese Oxides of Various Compositions in Sulfuric Acid Solutions Studied by Kinetic Methods, Inorganic Materials, 54 (2018) 66-71 10.1134/S002016851801003X.

[56] B. Babakhani, D.G. Ivey, Effect of electrodeposition conditions on the electrochemical capacitive behavior of synthesized manganese oxide electrodes, Journal of Power Sources, 196 (2011) 10762-10774 https://doi.org/10.1016/j.jpowsour.2011.08.102.

[57] M.I. Zaki, M.A. Hasan, L. Pasupulety, K. Kumari, Thermochemistry of manganese oxides in reactive gas atmospheres: Probing redox compositions in the decomposition course $MnO_2 \rightarrow MnO$, Thermochimica Acta, 303 (1997) 171-181 https://doi.org/10.1016/S0040-6031(97)00258-X.

[58] Z.B. Jildeh, J. Oberländer, P. Kirchner, P.H. Wagner, M.J. Schöning, Thermocatalytic Behavior of Manganese (IV) Oxide as Nanoporous Material on the Dissociation of a Gas Mixture Containing Hydrogen Peroxide, Nanomaterials (Basel), 8 (2018) 262 10.3390/nano8040262.

[59] M.R. Hoffmann, Kinetics and mechanism of oxidation of hydrogen sulfide by hydrogen peroxide in acidic solution, Environmental Science & Technology, 11 (1977) 61-66 10.1021/es60124a004.

[60] J. Herszage, M. dos Santos Afonso, Mechanism of Hydrogen Sulfide Oxidation by Manganese(IV) Oxide in Aqueous Solutions, Langmuir, 19 (2003) 9684-9692 10.1021/la034016p.

[61] J.E. Kostka, G.W. Luther, K.H. Nealson, Chemical and biological reduction of Mn (III)-pyrophosphate complexes: Potential importance of dissolved Mn (III) as an environmental oxidant, Geochimica et Cosmochimica Acta, 59 (1995) 885-894 https://doi.org/10.1016/0016-7037(95)00007-0.

[62] V.I. Kravtsov, Outer and inner sphere mechanims of electrochemical steps of the metal complexes electrode reactions, Journal of Electroanalytical Chemistry and Interfacial Electrochemistry, 69 (1976) 125-131 https://doi.org/10.1016/S0022-0728(76)80244-6.

[63] J. Melder, P. Bogdanoff, I. Zaharieva, S. Fiechter, H. Dau, P. Kurz, Water-Oxidation Electrocatalysis by Manganese Oxides: Syntheses, Electrode Preparations, Electrolytes and Two Fundamental Questions, Zeitschrift für Physikalische Chemie, 234 (2020) 925-978 doi:10.1515/zpch-2019-1491.

[64] H. Tamura, T. Oda, M. Nagayama, R. Furuichi, Acid-Base Dissociation of Surface Hydroxyl Groups on Manganese Dioxide in Aqueous Solutions, Journal of The Electrochemical Society, 136 (1989) 2782-2786 10.1149/1.2096286.

[65] X.H. Feng, L.M. Zhai, W.F. Tan, W. Zhao, F. Liu, J.Z. He, The controlling effect of pH on oxidation of Cr(III) by manganese oxide minerals, Journal of Colloid and Interface Science, 298 (2006) 258-266 https://doi.org/10.1016/j.jcis.2005.12.012.

[66] W. Yao, F.J. Millero, Adsorption of Phosphate on Manganese Dioxide in Seawater, Environmental Science & Technology, 30 (1996) 536-541 10.1021/es950290x.

[67] A.T. Stone, Reductive Dissolution of Manganese(III/Iv) Oxides by Substituted Phenols, Environmental Science & Technology, 21 (1987) 979-988 10.1021/es50001a011.

[68] A.T. Stone, J.J. Morgan, Reduction and dissolution of manganese(III) and manganese(IV) oxides by organics: 2. Survey of the reactivity of organics, Environmental Science & Technology, 18 (1984) 617-624 10.1021/es00126a010.

[69] P.K. Dutta, K. Rabaey, Z. Yuan, J. Keller, Spontaneous electrochemical removal of aqueous sulfide, Water Research, 42 (2008) 4965-4975 https://doi.org/10.1016/j.watres.2008.09.007.

[70] Y.S. Shih, J.L. Lee, Continuous solvent extraction of sulfur from the electrochemical oxidation of a basic sulfide solution in the CSTER system, Industrial & Engineering Chemistry Process Design and Development, 25 (1986) 834-836 10.1021/i200034a041.

[71] Yi Qing, Chen Qi, Zhang Ping, Electrochemical study of sulfide solution in the presence of surfactants, Journal of Environmental Sciences, 10  372-377

[72] A. Le Faou, B.S. Rajagopal, L. Daniels, G. Fauque, Thiosulfate, polythionates and elemental sulfur assimilation and reduction in the bacterial world, FEMS Microbiology Letters, 75 (1990) 351-382 https://doi.org/10.1016/0378-1097(90)90688-M.

**Supplementary Material**

**Manganese oxide coated TiO$_2$ nanotube-based electrode for efficient and selective electrocatalytic sulfide oxidation to colloidal sulfur**

*Natalia Sergienko[a,b], Jelena Radjenovic [a,c*]*

[a]*Catalan Institute for Water Research (ICRA), Scientific and Technological Park of the University of Girona, 17003 Girona, Spain*

[b] *University of Girona, Girona, Spain*

[c]*Catalan Institution for Research and Advanced Studies (ICREA), Passeig Lluís Companys 23, 08010 Barcelona, Spain*

*\* Corresponding author:*

*Jelena Radjenovic, Catalan Institute for Water Research (ICRA), c/Emili Grahit, 101, 17003 Girona, Spain*

Phone: + 34 972 18 33 80; Fax: +34 972 18 32 48; E-mail: jradjenovic@icra.cat

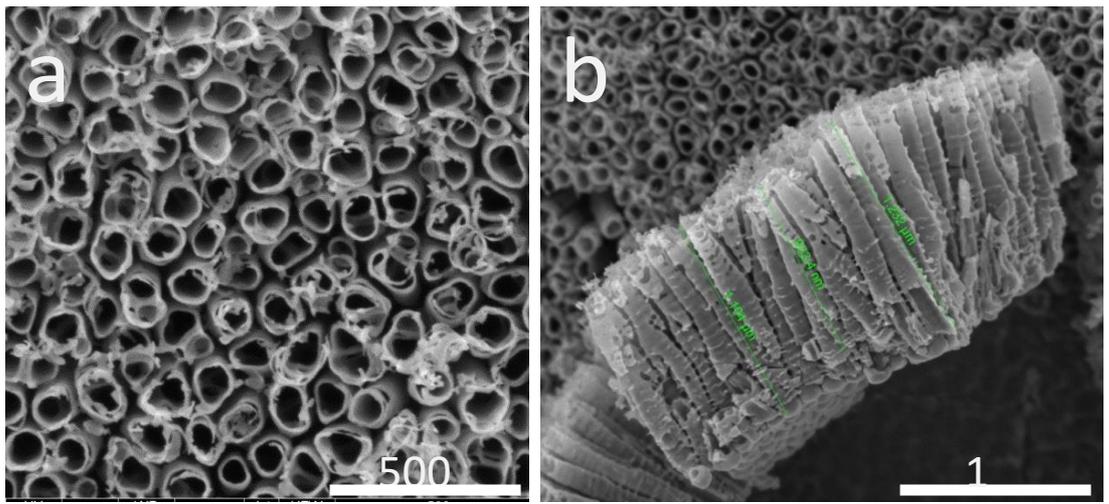

**Figure S1.** Top view and the cross section of the self-organized TiO$_2$ nanotube interlayer.

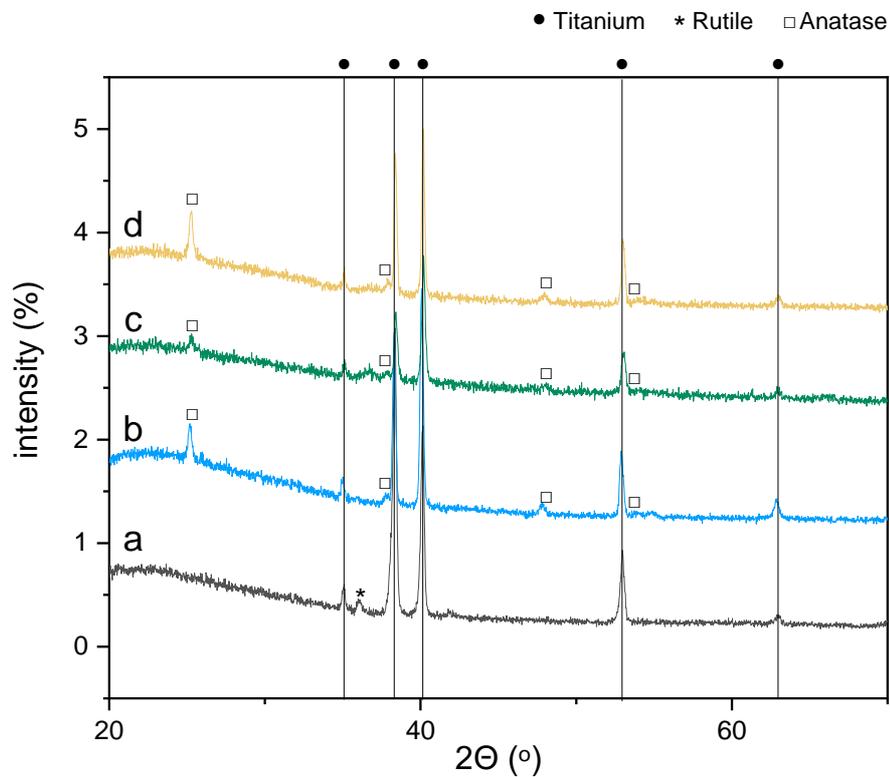

**Figure S2**. The XRD patterns of the: **a)** Ti plate, **b)** Ti plate with $TiO_2$ NTA interlayer, **c)** $TiO_2$ NTA coated with $Mn_xO_y$ in the presence of 0.5 M $H_2SO_4$, **d)** $TiO_2$ NTA coated with $Mn_xO_y$ under high acid concentration (e.g 0.5 M $H_2SO_4$), followed by the calcination step.

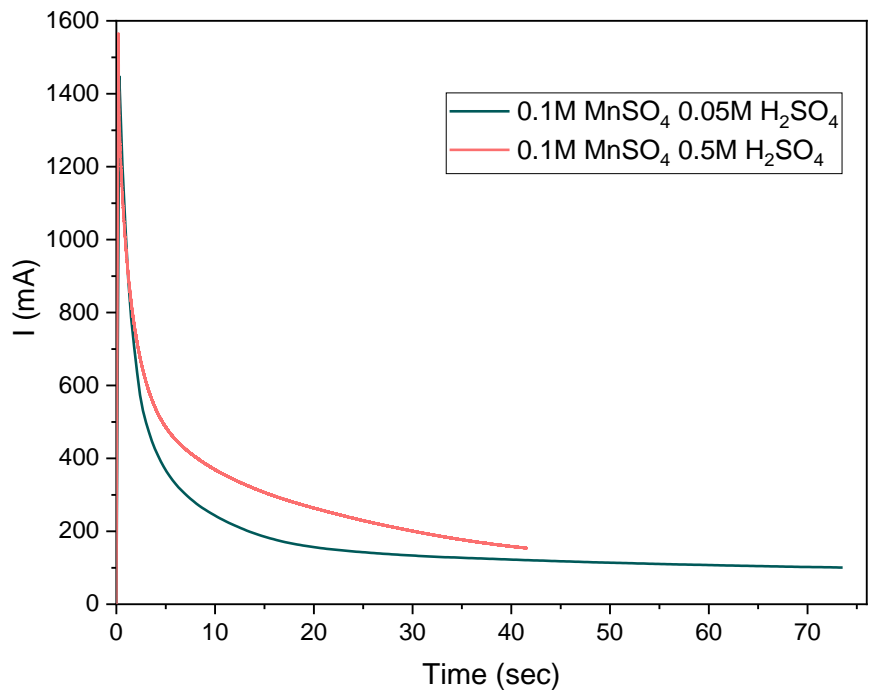

**Figure S3**. Chronoamperometries obtained during the $Mn_xO_y$ electrodeposition procedure performed in the electrolyte containing 0.1 M $MnSO_4$ and 0.05 M $H_2SO_4$ or 0.5 M $H_2SO_4$.

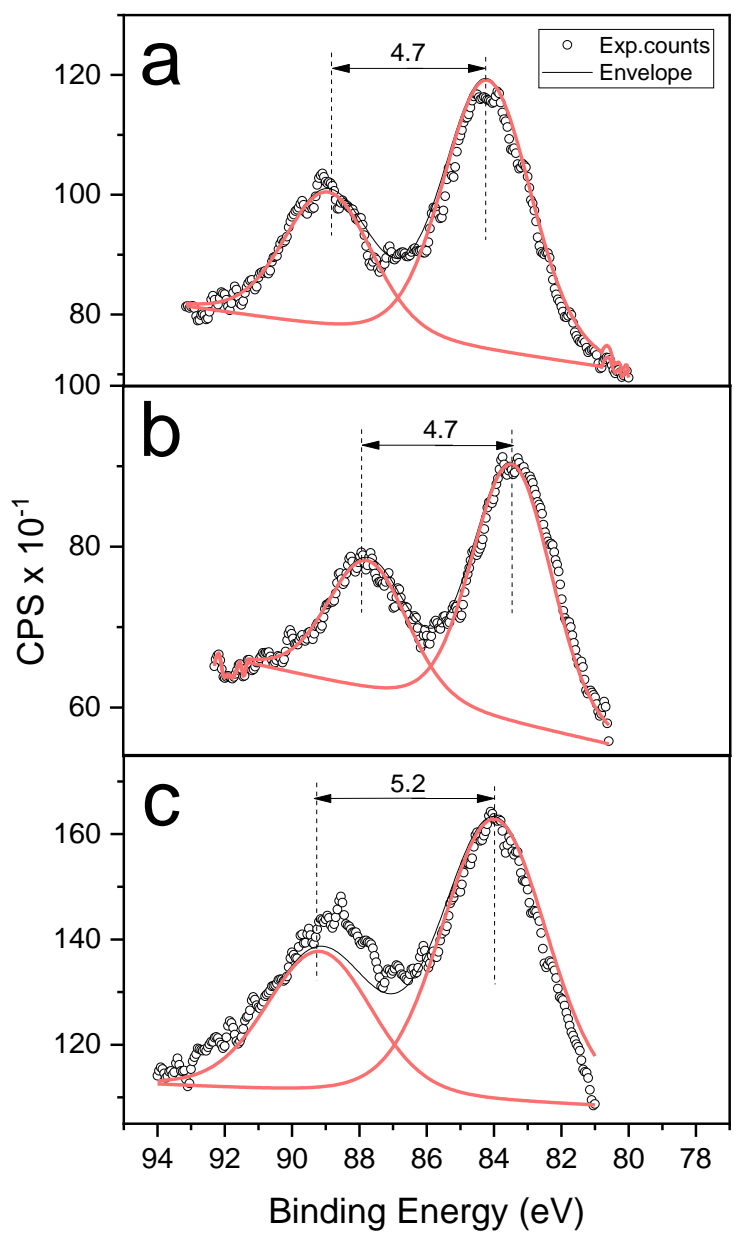

**Figure S4**. High resolution Mn3s XPS spectra of the TiO$_2$ NTA coated with Mn$_x$O$_y$ synthesized in the precursor solution containing 0.1 M MnSO$_4$ and **a)** 0.05 M H$_2$SO$_4$, **b)** 0.5 M H$_2$SO$_4$, and **c)** 0.5 M H$_2$SO$_4$, followed by the calcination step.

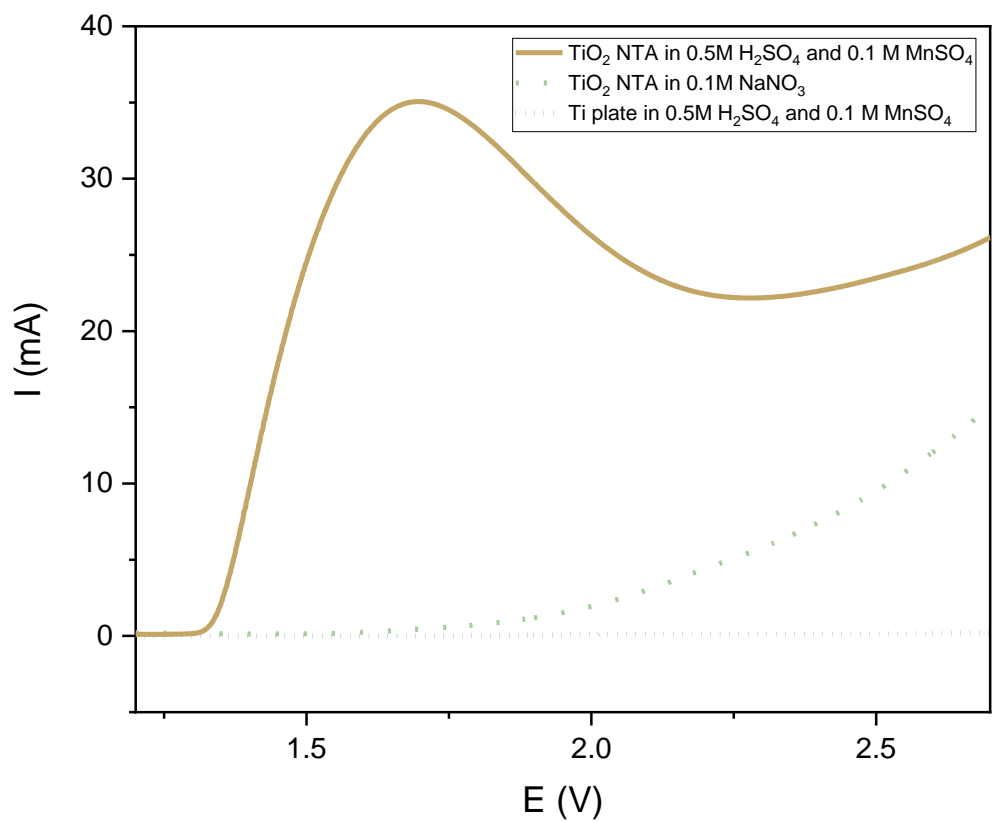

2  **Figure S5.** Linear sweep voltammograms performed using TiO$_2$ NTA in 0.5 M H$_2$SO$_4$
3  and 0.1 M MnSO$_4$ or 0.1 M NaNO$_3$ and Ti plate in 0.5 M H$_2$SO$_4$ and 0.1 M MnSO$_4$.

4
5
6
7
8
9
10
11
12

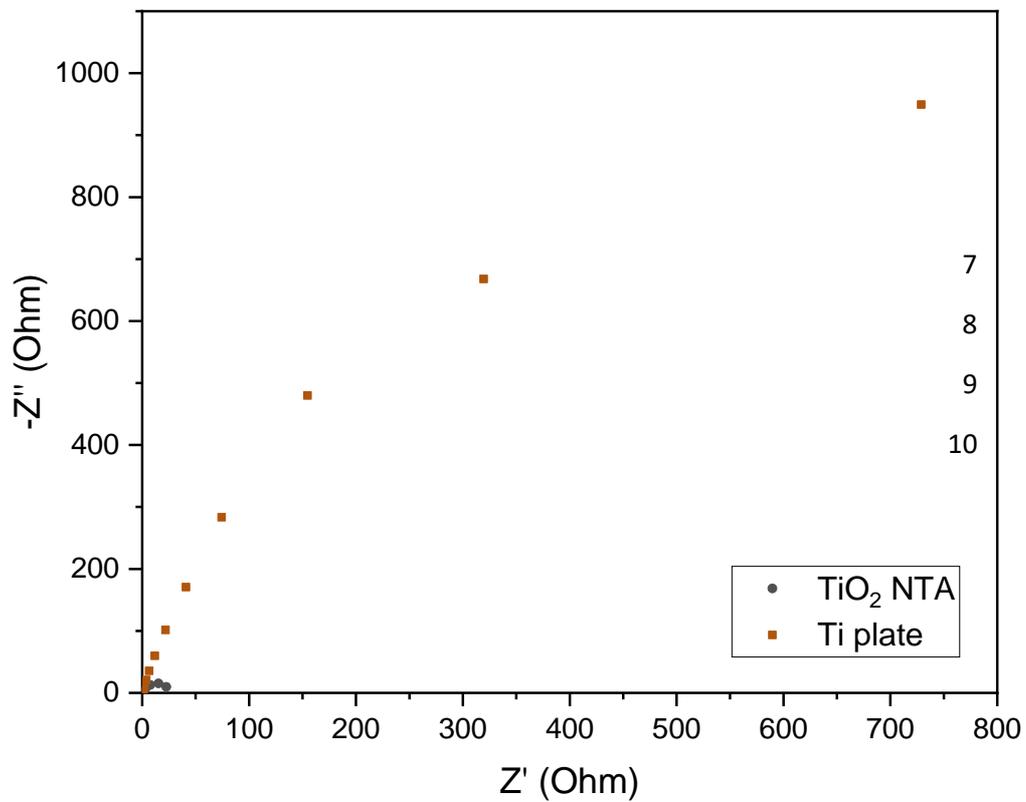

**Figure S6**. Nyquist plots of the EIS measurements at 3.2 V of Ti plate and Ti plate with TiO$_2$ NTA interlayer.

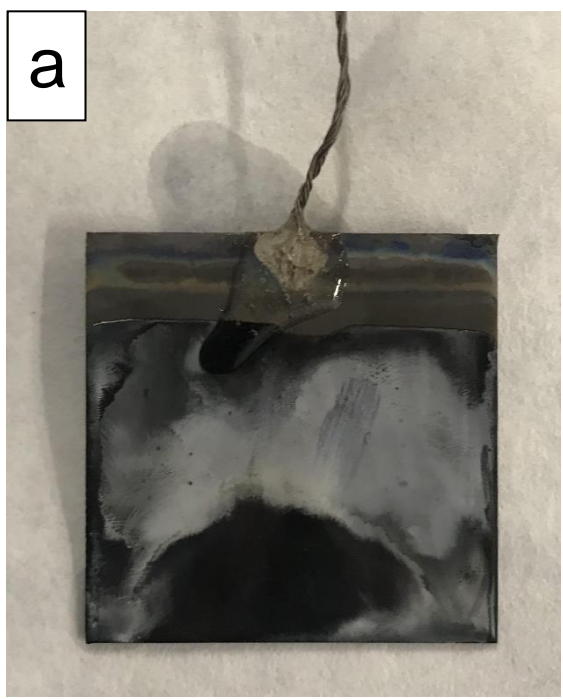 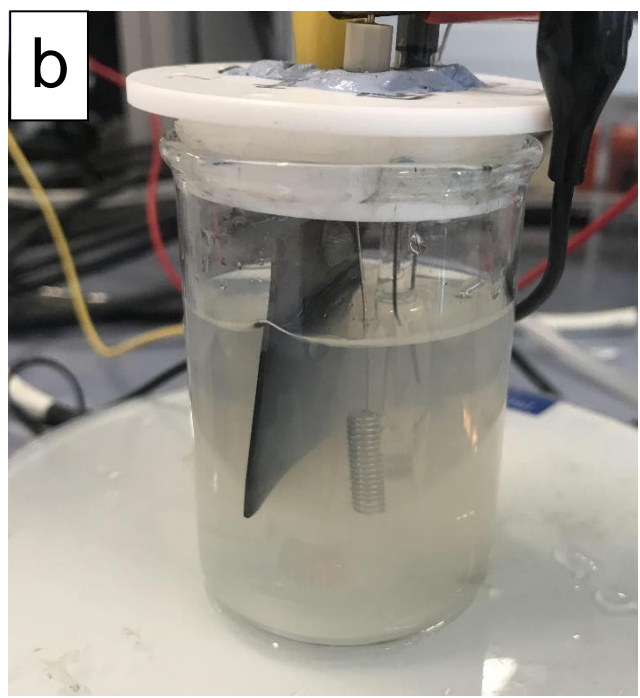

**Figure S7**. Images of the: **a)** Ti/TiO$_2$NTA-MnO$_2$ electrode after the electrochemical sulfide removal test performed at pH 12, **b)** colloidal sulfur solution produced during electrochemical sulfide removal using the Ti/TiO$_2$NTA-Mn$_x$O$_y$ electrodes at pH 8.

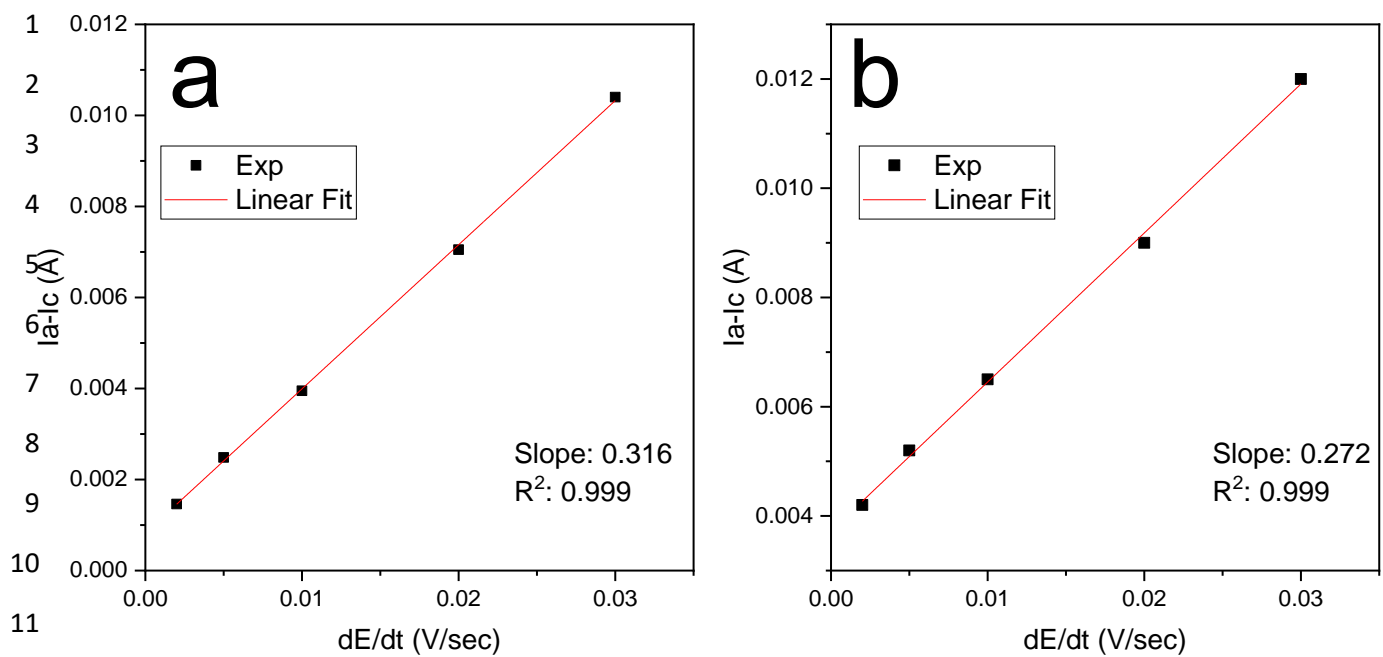

**Figure S8**. Determination of the double-layer capacitance at scan rates between 30 and 2 mV s$^{-1}$ by performing CV in 0.1 M NaNO$_3$ electrolyte. The plots of charging currents versus the scan rates of the: **a)** Ti/TiO$_2$ NTA-MnO$_2$ electrodes synthesized in the presence of 0.5 M H$_2$SO$_4$, **b)** Ti/TiO$_2$ NTA-MnO$_2$ electrodes synthesized in the presence of 0.05 M H$_2$SO$_4$.

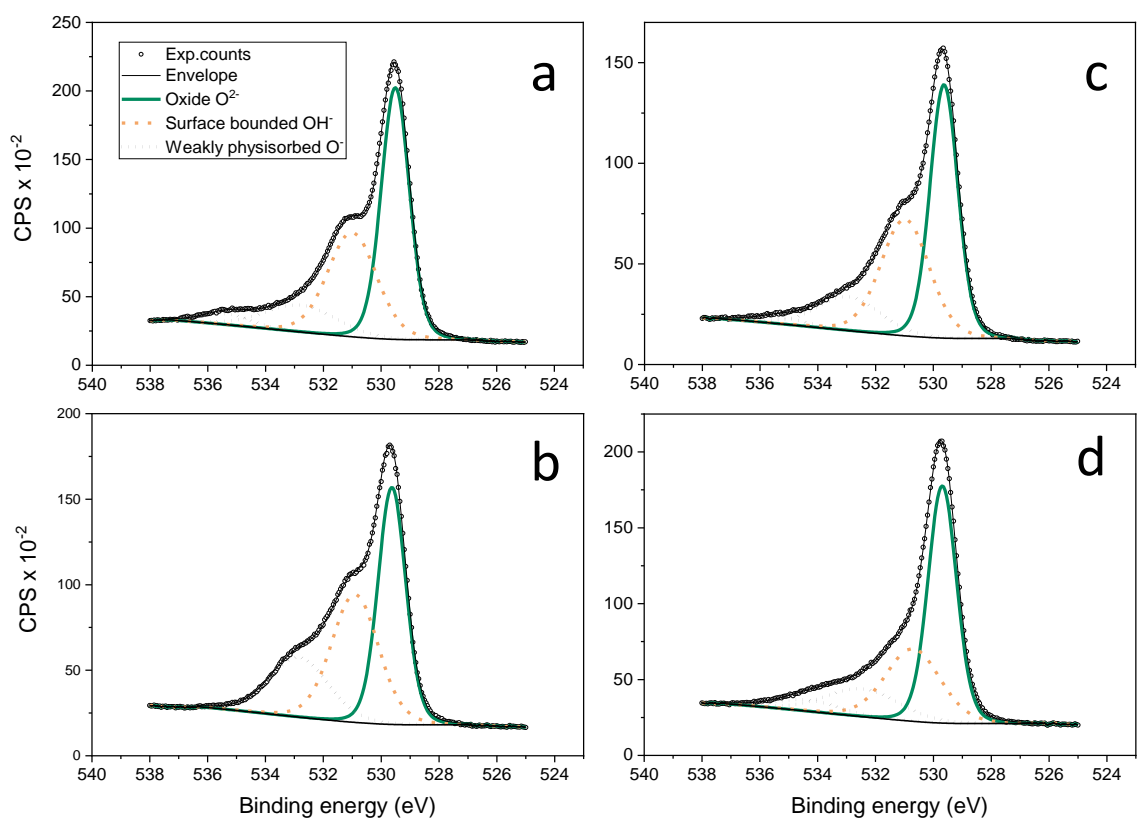

**Figure S9**. O 1s photoelectron spectra for **a)** Ti/TiO$_2$ NTA/MnO$_2$ sample prior to the sulfide removal test at pH 8, **b)** Ti/TiO$_2$ NTA/MnO$_2$ sample after the sulfide removal test at pH 8, **c)** Ti/TiO$_2$ NTA/MnO$_2$ sample prior to the sulfide removal test at pH 12, **d)** Ti/TiO$_2$ NTA/MnO$_2$ sample after the sulfide removal test at pH 12.

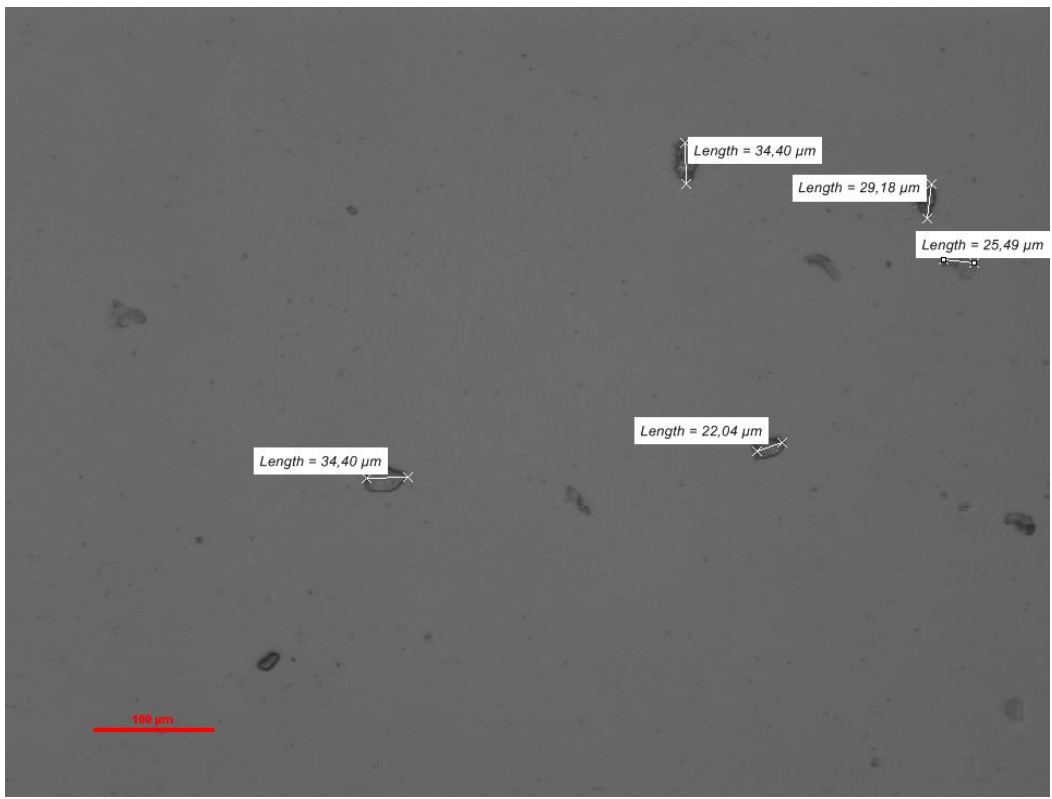

**Figure S10**. Microscopic image of the sulfur particles produced during electrochemical sulfide removal using Ti/TiO$_2$NTA-Mn$_x$O$_y$ electrodes at pH 8.

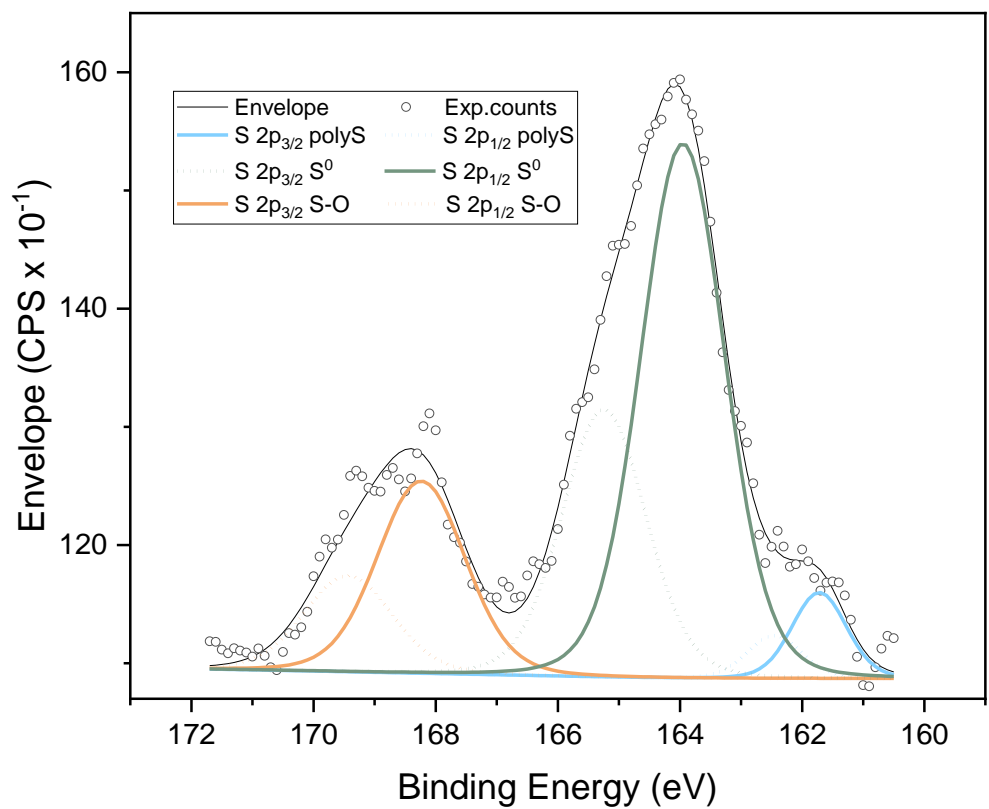

**Figure S11**. High resolution S2p XPS spectra of the Ti/TiO$_2$NTA-Mn$_x$O$_y$ electrode after the electrochemical sulfide removal test performed at pH 12.

**Table S1**. First-order sulfide removal rates (h$^{-1}$) and the total dissolved manganese measured in the open circuit experiments, and in the chronoamperometric experiments at 0.4 - 0.8 V/SHE applied to the Ti/TiO$_2$NTA-MnO$_2$ anode using 2.6 mM NaNO$_3$ and 2 mM HS$^-$ containing electrolyte at pH 12.

|  | First order removal rate, h$^{-1}$ | Total dissolved Mn, mg/L |
|---|---|---|
| **OC** | 0.13±0.01 | 15.03 |
| **at 0.4 V** | 0.21±0.07 | 0.7 |
| **at 0.6 V** | 0.52±0.1 | 0.18 |
| **at 0.8 V** | 0.98±0.2 | 0.08 |

**Table S2.** First-order sulfide removal rates (h$^{-1}$) and the total dissolved manganese measured in the chronoamperometric experiments at 0.8 V/SHE applied to the Ti/TiO$_2$NTA-MnO$_2$ anode using 2.6 mM NaNO$_3$ and 0.9 – 3.2 mM HS$^-$ containing electrolyte at pH 12.

|  | First order removal rate, h$^{-1}$ | Total dissolved Mn, mg/L |
|---|---|---|
| **at 0.9 mM** | 1.04±0.05 | 0.138 |
| **at 2 mM** | 0.98±0.2 | 0.08 |
| **at 3.2 mM** | 0.62±0.1 | 0.196 |

**Table S3**. First-order sulfide removal rates (h$^{-1}$) and the total dissolved manganese in the chronoamperometric experiments at 0.8 V/SHE applied to the TI/TiO$_2$NTA-MnO$_2$ anode using 2.6 mM NaNO$_3$ and 2 mM HS$^-$ containing electrolyte at pH 8 and pH 12.

|  | First order removal rate, h$^{-1}$ | Total dissolved Mn, mg/L |
|---|---|---|
| **at pH 8** | 1.32±0.2 | 0.114 |
| **at pH 12** | 0.98±0.2 | 0.08 |

Table S4. First-order sulfide removal rates (h$^{-1}$) in the repeated chronoamperometric experiments at 0.8 V/SHE applied to the Ti/TiO$_2$NTA-MnO$_2$ anode using **a)** 2.6 mM NaNO$_3$ and 2 mM HS$^-$ containing electrolyte at pH 12, **b)** 2.6 mM NaNO$_3$ and 2 mM HS$^-$ containing electrolyte at pH 8, **c)** real sewage amended with 2 mM HS$^-$ containing electrolyte at pH 8.

|  | First order removal rate (a), h$^{-1}$ | Eeo (a), W h L$^{-1}$ | First order removal rate (b), h$^{-1}$ | Eeo (b), W h L$^{-1}$ | First order removal rate (c), h$^{-1}$ | Eeo (c), W h L$^{-1}$ |
|---|---|---|---|---|---|---|
| **1$^{st}$ application** | 0.81±0.03 | 0.09 | 1.01±0.04 | 0.54 | 0.68±0.06 | 0.29 |
| **2$^{nd}$ application** | 0.41±0.06 | 0.18 | 0.96±0.06 | 0.44 | 0.7±0.065 | 0.26 |
| **3$^{d}$ application** | 0.19±0.08 | 0.79 | 1.06±0.1 | 0.42 | 0.69±0.09 | 0.31 |